\documentclass[
    aps,
    prd,
    reprint,
    superscriptaddress,
    nolongbibliography,
    nobibnotes,
    nofootinbib,
]{revtex4-2}

\usepackage[dvipsnames]{xcolor}
\definecolor{linkcolor}{rgb}{0.1, 0.5, 0.7}

\usepackage[
    colorlinks=true,
    linkcolor=linkcolor,
    citecolor=linkcolor,
    filecolor=linkcolor,
    urlcolor=linkcolor,
    pdfusetitle,
]{hyperref}

\usepackage{amssymb}
\usepackage{amsmath}
\usepackage{physics}
\usepackage{orcidlink}

\newcommand{\comment}[1]{}

\newcommand{\notts}{\affiliation{Nottingham Centre of Gravity \& School of Mathematical Sciences, University of Nottingham,\\University Park, Nottingham, NG7 2RD, United Kingdom}}
\newcommand{\ligo}{\affiliation{LIGO Laboratory, Massachusetts Institute of Technology, Cambridge, MA 02139, USA}}
\newcommand{\mki}{\affiliation{Kavli Institute for Astrophysics and Space Research, Massachusetts Institute of Technology, Cambridge, MA 02139, USA}}
\renewcommand{\mit}{\affiliation{Department of Physics, Massachusetts Institute of Technology, Cambridge, MA 02139, USA}}

\begin{document}

\title{Measurement prospects for the pair-instability mass cutoff with gravitational waves}

\author{Matthew Mould\,\orcidlink{0000-0001-5460-2910}}
\email{matthew.mould@nottingham.ac.uk}
\notts\ligo\mki\mit

\author{Jack Heinzel\,\orcidlink{0000-0002-5794-821X}}
\ligo\mki\mit

\author{Sofía Álvarez-López\,\orcidlink{0009-0003-8040-4936}}
\ligo\mki\mit

\author{Cailin Plunkett\,\orcidlink{0000-0002-1144-6708}}
\ligo\mki\mit

\author{Noah E. Wolfe\,\orcidlink{0000-0003-2540-3845}}
\ligo\mki\mit

\author{Salvatore Vitale\,\orcidlink{0000-0003-2700-0767}}
\ligo\mki\mit

\date{\today}

\begin{abstract}
Pair-instability supernovae leave behind no compact remnants, resulting in a predicted gap in the distribution of stellar black-hole masses. Gravitational waves from binary black-hole mergers probe the relevant mass range and analyses of the LIGO--Virgo--KAGRA catalog (GWTC-4) indicate a possible mass cutoff at 40--50$M_\odot$. However, the robustness of this result remains unclear. To this end, using full Bayesian parameter estimation, we simulate gravitational-wave catalogs with and without such a mass cutoff, then test whether its presence or absence is correctly inferred with parametric population models. For catalogs similar to GWTC-4, confident identification of a cutoff is not guaranteed, but the best constraints among our simulations are compatible with results from GWTC-4 when the model includes a cutoff. Conversely, spurious identification of a cutoff is unlikely. For catalogs expected by the end of the O4 observing run, uncertainty in the cutoff mass is reduced by $\gtrsim20\%$, but a cutoff at 40--50$M_\odot$ yields only a lower bound on the $^{12}\mathrm{C}(\alpha,\gamma)^{16}\mathrm{O}$ reaction rate, our most stringent constraints on the S-factor at 300\,keV being $S_{300}\gtrsim125$\,keV\,b at 90\% credibility. Relative uncertainties on the Hubble parameter $H_0$ from gravitational-wave data alone can still be up to 100\%. We also analyze GWTC-4 with the nonparametric \textsc{PixelPop} population model, finding that some mass features are more prominent than in parametric models but a sharp cutoff is not required. However, the parametric model passes a likelihood-based predictive test in GWTC-4 and the \textsc{PixelPop} results are consistent with those from our simulated catalogs with a cutoff. Such tests are necessary to make astrophysical claims from gravitational-wave catalogs.
\end{abstract}

\maketitle

\section{Introduction}
\label{section: Introduction}

The LIGO \cite{LIGOScientific:2014pky, Capote:2024rmo, LIGO:2024kkz}, Virgo \cite{VIRGO:2014yos}, and KAGRA \cite{KAGRA:2020tym} (LVK) gravitational-wave (GW) detectors have made $>100$ observations of the mergers of stellar-mass black holes (BHs) in compact binaries \cite{LIGOScientific:2018mvr, LIGOScientific:2020ibl, LIGOScientific:2021usb, KAGRA:2021vkt, LIGOScientific:2025slb}. A fundamental open question is the origin of these systems \cite{Mapelli:2020vfa, Mapelli2020, Mandel:2021smh}, but astrophysical insights have been tentative \cite{LIGOScientific:2018jsj, LIGOScientific:2020kqk, KAGRA:2021duu, LIGOScientific:2025pvj}. One particular physical process that may occur in the late-stage evolution of a massive progenitor star is the pair instability \cite{Fowler:1964zz, Barkat:1967zz, 1967ApJ...148..803R, fraley1968supernovae}. Core contraction increases the temperature sufficiently for pair production, reducing internal pressure support. Rebounding shocks powered by explosive nuclear burning can be so energetic as to completely disrupt the star in a pair-instability supernova (PISN), or eject material in a series of pulsational episodes with eventual collapse to a BH in a pulsational PISN \cite{Woosley:2016hmi}. The outcome is a ``mass gap'' from $\sim50$--120$M_\odot$ in which BHs do not form directly via stellar collapse \cite{Belczynski:2016jno, Stevenson_2019, Farmer:2020xne, Mehta:2021fgz, Woosley:2021xba}. Though there are candidate PISN observations, these are not confirmed (see, e.g., Refs.~\cite{Hendriks:2023yrw, Gabrielli:2024rub} and references therein). There are also significant theoretical uncertainties that impact the location of the purported mass gap \cite{Farmer:2019jed, Mapelli:2019ipt, Renzo:2020rzx, Marchant:2020haw, Costa:2020xbc, Woosley:2021xba}---most notably, nuclear reaction rates for $^{12}\mathrm{C}(\alpha,\gamma)^{16}\mathrm{O}$ carbon--oxygen burning \cite{deBoer:2017ldl, Takahashi:2018kkb, Farmer:2020xne, Mehta:2021fgz, Farag:2022jcc}.

As GW observations constrain the source BH properties such as mass, they can be used to search for the imprints of (pulsational) PISNe. Following the first GW transient catalog (GWTC-1) \cite{LIGOScientific:2018mvr}, Ref.~\cite{LIGOScientific:2018jsj} found that $<1\%$ of sources in the underlying population have masses $>45M_\odot$, potentially consistent with a PISN gap. After GWTC-2 \cite{LIGOScientific:2020ibl}, Ref.~\cite{LIGOScientific:2020kqk} found that---while a sharp mass cutoff is not required to fit the data---only $\sim3\%$ of systems have component masses $>45M_\odot$. They also found evidence of an overdensity of sources with mass $\sim35M_\odot$, possibly due to pulsational PISNe \cite{Talbot:2018cva}; later work suggested this interpretation is disfavored or offered alternative explanations \cite{Antonini:2022vib, Hendriks:2023yrw, Golomb:2023vxm, Roy:2025ktr, Sridhar:2025kvi}. Ref.~\cite{KAGRA:2021duu} analyzed the population following the additions of events through GWTC-3 \cite{LIGOScientific:2021usb, KAGRA:2021vkt}, finding that a strongly suppressed merger rate in the PISN mass gap is not required by the GW data, reflecting the addition of higher-mass detections.

\begin{table*}
\caption{Summary of the mass population models we consider, along with the Bayes factors $\mathcal{B}$ in favor of each phenomenological model (top three rows) relative to that which mostly closely resembles the default model of Ref.~\cite{LIGOScientific:2025pvj}, \textsc{Broken Power Law + 2 Peaks} (third row). We do not compute the Bayesian evidence for \textsc{PixelPop} as comparison between nonparametric and parametric models are not meaningful. Bayes factors are from analyses of the binary BH mergers through GWTC-4 but excluding GW231123, while those from GWTC-3 are in brackets.}
\centering
\setlength{\tabcolsep}{5pt}
\begin{tabular}{lll}
\hline\hline
Model & Description & $\log_{10}\mathcal{B}$ \\
\hline\hline
\textsc{Single Power Law + 2 Peaks} & $m_1$ distribution consists of a power law and two Gaussians, & $2.8\pm0.2$ ($0.3\pm0.1$) \\
\textsc{+ Cutoff} & with low-mass smoothing, free minimum and maximum. & \\
& $m_2$ distribution is a power law with the same minimum and & \\
& low-mass smoothing as $m_1$, but an independent maximum. & \\[5pt]
\textsc{Single Power Law + 2 Peaks} & As above, but $m_1$ and $m_2$ have the same maximum. & $0.5\pm0.2$ ($-0.1\pm0.1$) \\[5pt]
\textsc{Broken Power Law + 2 Peaks} & As above, but with a broken power law in $m_1$. & 0.0 (0.0) \\[5pt]
\textsc{PixelPop} & Binned $m_1$--$m_2$ plane with nearest-neighbor coupling. & --- \\
\hline\hline
\end{tabular}
\label{table: models}
\end{table*}

Independent analyses did find evidence for a PISN cutoff, however, starting from GWTC-2 \cite{Wang:2021mdt, Baxter:2021swn}. Ref.~\cite{Mould:2022ccw} then reanalyzed GWTC-3 with a simulation-based population model accounting for both stellar-origin BHs and repeated BH mergers \cite{Gerosa:2021mno}, finding a lower edge for the PISN mass gap $\sim40M_\odot$ and a positive population-level correlation between BH masses and spins. Subsequent work supported these findings with other physically motivated parametrizations \cite{Karathanasis:2022rtr, Golomb:2023vxm} and more agnostic population models \cite{Wang:2022gnx, Li:2023yyt, Pierra:2024fbl, Antonini:2024het, Antonini:2025zzw, MaganaHernandez:2025fkm}.

The recent release of GWTC-4 \cite{LIGOScientific:2025slb} more than doubles the number of confident binary BH detections compared to GWTC-3, allowing for sharper constraints on the population \cite{LIGOScientific:2025pvj}. Refs.~\cite{Tong:2025wpz, Guttman:2025jkv} found evidence for a truncation or gap starting at $\sim45M_\odot$ in the mass distribution of the lighter of the two BHs in merging binaries (the secondary), but not in that of the heavier (primary) BH, suggesting an astrophysical mechanism that populates the PISN gap for the preferentially heavier BH. The source of GW231123 sits above the gap as the heaviest BH merger in the LVK catalog \cite{LIGOScientific:2025rsn}, which---if its component BHs are the direct result of stellar collapse---implies the gap ends at $\sim120M_\odot$. Whereas the lower edge of the gap is informed by the entire catalog, the upper edge depends on strong prior assumptions and ambiguous interpretation of GW231123 alone \cite{Mandel:2025qnh, Tenorio:2026dcc}: one can either constrain the upper edge of the PISN gap \textit{assuming} the BHs of GW231123 trace stellar collapse, or constrain the probability that this binary requires an alternative formation scenario \textit{assuming} a gap location \cite{LIGOScientific:2025rsn}, but not both. Ref.~\cite{Antonini:2025ilj} analyzed GWTC-4 using a model with a transition in the BH spin distribution as a function of mass \cite{Antonini:2024het, Antonini:2025zzw} rather than a separate secondary-mass truncation, but similarly found a feature at $\sim45M_\odot$. In these works, associating the inferred mass feature with PISNe placed constraints on the $^{12}\mathrm{C}(\alpha,\gamma)^{16}\mathrm{O}$ reaction rate \cite{Farmer:2020xne}, albeit with uncertainties $\gtrsim100\%$; cf. $\approx20\%$ combined uncertainty from nuclear theory and laboratory experiments extrapolated to temperatures relevant for nuclear burning in massive stars \cite{deBoer:2017ldl, Shen:2023rco, Mukhamedzhanov:2025uvy}.

The growing GW dataset and agreement between subsequent population analyses \cite{Banagiri:2025dmy, Tong:2025xir, Plunkett:2026pxt, Farah:2026jlc, Vijaykumar:2026zjy} lend credence to the (re)emerging interpretation that PISNe have a detectable and \textit{detected} impact on the BH mass spectrum; however, other recent works disagree on this interpretation \cite{Wang:2025nhf, Ray:2025xti}. Overall, no analyses have tested the robustness of these conclusions, i.e., \textit{``assuming such a feature really is present in the population, how well should we expect to measure it?''}, and conversely, \textit{``is it possible to spuriously infer such a feature when it does not exist?''}. Unfortunately, there is a general lack of such validation studies for GW population results at present (though see, e.g., Refs.~\cite{Farah:2023vsc, Toubiana:2023egi, Miller:2024sui, Biscoveanu:2025jpc, Vitale:2025lms} for a few examples), without which any astrophysical conclusions are not solid. We tackle this issue here in the context of measuring the PISN mass cutoff from GW catalogs, serving (1) as an example of how population measurements should be tested going forward, and (2) as a projection of how well PISN physics (such as the edge of the mass gap and nuclear reaction rates) as well as cosmological parameters may be constrained by the end of the fourth LVK observing run (O4).

In Section~\ref{section: Observational summary}, we analyze the latest GW dataset and summarize population-level constraints, including the purported pair-instability mass cutoff. We perform a comprehensive suite of simulation studies to validate the results of these analyses in Section~\ref{section: Validating GWTC-4 constraints} and make projections for future catalogs in Section~\ref{section: Projections for the end of O4}. Concluding discussion is presented in Section~\ref{section: Conclusions}.

\section{Observational summary}
\label{section: Observational summary}

\begin{figure*}
\centering
\includegraphics[width=0.9\textwidth]{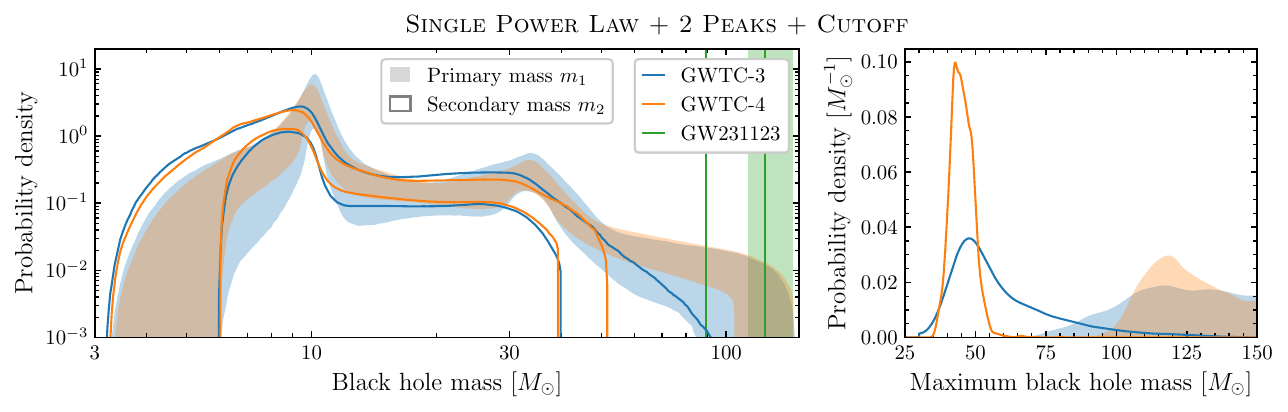}\vspace{-5pt}
\includegraphics[width=0.9\textwidth]{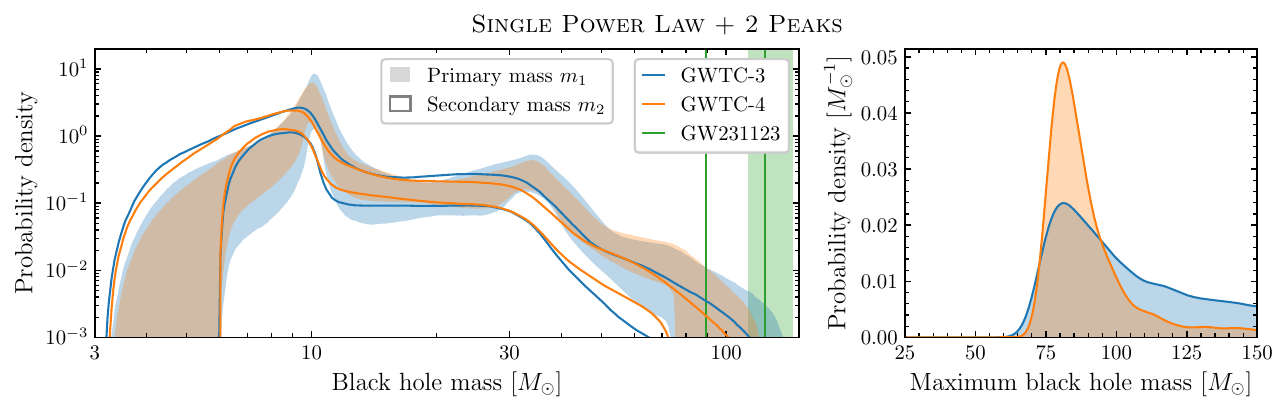}\vspace{-5pt}
\includegraphics[width=0.9\textwidth]{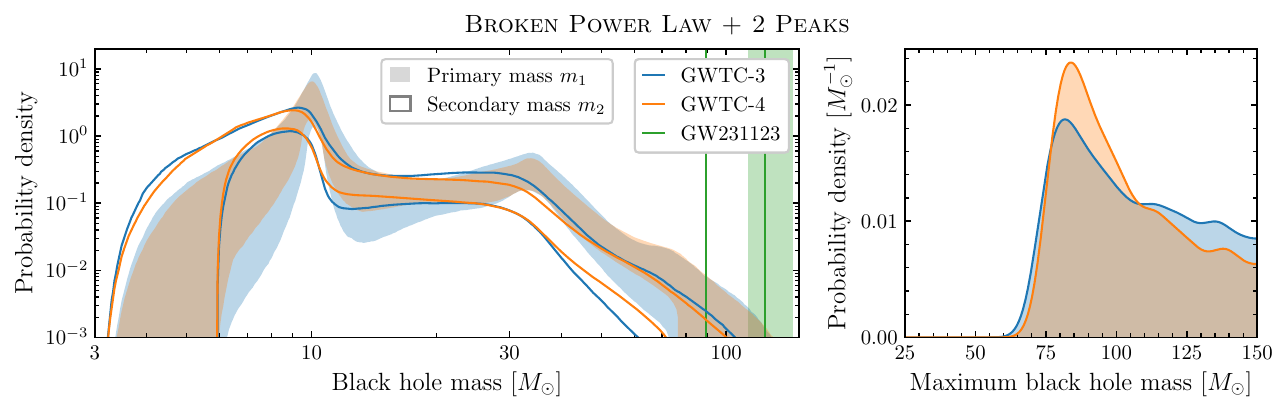}\vspace{-5pt}
\includegraphics[width=0.9\textwidth]{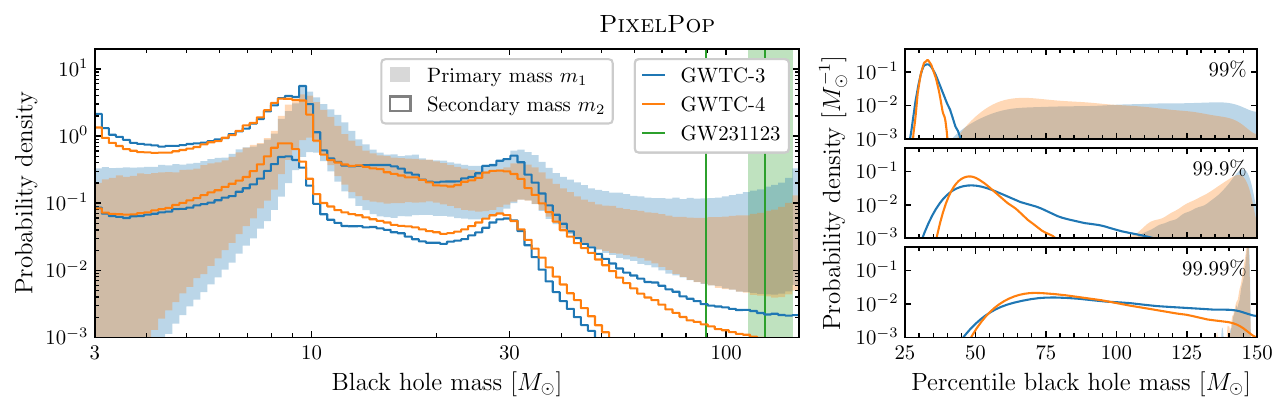}
\caption{Mass distributions of BHs in merging binaries inferred from GWTC-3 (blue) and GWTC-4 excluding GW231123 (orange) using three parametric models (top three rows) and a nonparametric model (bottom row). The masses for GW231123 (green) are included for reference. The left column shows central 90\% posterior credible regions, with population constraints in terms of the probability density of logarithmic BH mass. The right column shows the posteriors of the maximum-mass parameters for the parametric models and three upper mass percentiles (99\%, 99.9\%, and 99.99\%) for the nonparametric model. Primary ($m_1$) and secondary ($m_2$) BH masses are distinguished by filled and unfilled regions, respectively.}
\label{figure: observations}
\end{figure*}

\subsection{Data and models}
\label{section: Data and models}

We summarize results from our reanalyses of the cumulative GWTC-4 catalog \cite{LIGOScientific:2025slb, LIGOScientific:2025snk} of GW detections confidently identified as binary BH mergers, following Ref.~\cite{LIGOScientific:2025pvj}. Unlike Ref.~\cite{Tong:2025wpz} who inferred both lower and upper mass-gap edges, we exclude GW231123 from our analyses owing to its uncertain interpretation and focus only on the lower edge. For comparison, we also analyze the 69 confident BH mergers through GWTC-3 \cite{KAGRA:2021duu}.

We consider three phenomenological population models similar to but not the same as those in Refs.~\cite{LIGOScientific:2025pvj, Tong:2025wpz}. These are described in Appendix~\ref{section: Models} and summarized in Table~\ref{table: models}, including Bayesian evidences from GWTC-4 and GWTC-3. We also use the more flexible \textsc{PixelPop} model \cite{Heinzel:2024jlc, Heinzel:2024hva, Alvarez-Lopez:2025ltt} to test the impact of model assumptions. Though we infer the population over binary BH masses, spins, and merger redshift, here we focus on BH masses, as relevant for the PISN cutoff. The BH mass distributions inferred with each model are shown in Figure~\ref{figure: observations}.

\subsection{Parametric results}
\label{section: Parametric results}

Our phenomenological-model results confirm those of Ref.~\cite{Tong:2025wpz}. The model that allows for a truncation in secondary masses $m_2$ different from that in primary masses $m_1$---\textsc{Single Power Law + 2 Peaks + Cutoff}---is strongly preferred, according to Bayes factors $\log_{10}\mathcal{B}\approx2.3$ over the same model but with equal $m_1$ and $m_2$ maxima---\textsc{Single Power Law + 2 Peaks}---and $\log_{10}\mathcal{B}\approx2.8$ over our model most similar to the default from Ref.~\cite{LIGOScientific:2025pvj}---\textsc{Broken Power Law + 2 Peaks}. This evidence becomes significant only after the addition of new observations in GWTC-4, as there is no model preference according to Bayes factors within numerical uncertainties based on GWTC-3 alone.

As seen in Figure~\ref{figure: observations}, the preferred model (top row) finds posterior median and 90\% equal-tailed credible interval (CI) for the maximum $m_2$ of $45_{-5}^{+7}M_\odot$ from GWTC-4, much more constrained than $53_{-13}^{+46}M_\odot$ from GWTC-3. The maximum $m_1$ is not constrained as well, but is $>100M_\odot$ with 99\% credibility. This is in stark contrast to the two models (second and third rows of Figure~\ref{figure: observations}) that assume equal maxima for $m_1$ and $m_2$. For example, for the \textsc{Single Power Law + 2 Peaks} model, the posterior for the maximum mass is peaked more strongly at $85_{-9}^{+36}M_\odot$ compared to the GWTC-3 measurement of $94_{-20}^{+47}M_\odot$, though still with a tail to higher masses in the posterior. The discrepant maxima between models highlight the impact and potential misspecification of prior model choices \cite{Mandel:2025qnh}. Though we excluded GW231123, this does not impact the inferred $m_2$ cutoff or the broader mass distribution (cf. Ref.~\cite{Tong:2025wpz}).

The \textsc{Broken Power Law + 2 Peaks} model is only preferred (according to the Bayes factor) over the \textsc{Single Power Law + 2 Peaks} model when the $m_1$ and $m_2$ maxima are fixed as in Ref.~\cite{LIGOScientific:2025pvj} (not shown in Table~\ref{table: models}). When a broken power law is included, the power law is steeper after the break at only 88\% (78\%) credibility for GWTC-4 (GWTC-3). Its location is not well constrained from the prior and the measurement uncertainty does not reduce from GWTC-3 to GWTC-4; cf. a $\sim30\%$ reduction in uncertainty on the location of the peak of the mass distribution at $\sim10M_\odot$ between catalogs. Allowing for a broken power law \textit{and} independent $m_1$ and $m_2$ maxima yields a Bayes factor ($\log_{10}\mathcal{B}\approx2.4\pm0.2$, also not shown in Table~\ref{table: models}) lower than the equivalent model with an unbroken power law (\textsc{Single Power Law + 2 Peaks + Cutoff}).

Assuming the $m_2$ maximum in the \textsc{Single Power Law + 2 Peaks + Cutoff} model reflects the lower edge of the PISN mass gap, we follow Refs.~\cite{Farmer:2020xne, farmer_2020_4281044} to convert its posterior (but limit the BH mass range to 40--100$M_\odot$ to avoid errors from extrapolating their simulations, excluding $\approx8\%$ of posterior samples) to that for the astrophysical S-factor $S_{300}$ for the $^{12}\mathrm{C}(\alpha,\gamma)^{16}\mathrm{O}$ nuclear reaction at a temperature 300\,keV---a cross section with significant influence on post main sequence stellar evolution and the pair-instability gap \cite{deBoer:2017ldl}. We find $S_{300}=233_{-108}^{+193}$\,keV\,b, in broad agreement with Refs.~\cite{Tong:2025wpz, Antonini:2025ilj}. Because of the large uncertainty, this is also broadly consistent with independent constraints \cite{deBoer:2017ldl, Shen:2023rco, Mukhamedzhanov:2025uvy}, though Refs.~\cite{Hendriks:2023yrw, Wang:2025nhf} suggest higher values for the maximum BH mass produced by pulsational PISNe may be more likely.

We also test whether or not the identification of an $m_2$ mass cutoff improves cosmological constraints based only on GW data using the so-called ``spectral-siren'' method \cite{Taylor:2011fs, LIGOScientific:2025jau}. Following Ref.~\cite{LIGOScientific:2025jau}, we assume a flat $\Lambda$CDM model
with matter density fixed to $\Omega_m=0.3075$ \cite{Planck:2015fie}
and infer the Hubble parameter $H_0$. For the parametric models in Table~\ref{table: models}, we find that the widths of the 90\% posterior CIs are $\sim 60 \, (80) \, \mathrm{km} \, \mathrm{Mpc}^{-1} \, \mathrm{s}^{-1}$ for GWTC-4 (GWTC-3). Models allowing for the $m_2$ mass cutoff do have the smallest uncertainties, but only by $\sim10\%$.

\subsection{Nonparametric results}
\label{section: Nonparametric results}

Next, we turn to the nonparametric constraints in the bottom row of Figure~\ref{figure: observations}, using \textsc{PixelPop} to infer the BH merger rate in $m_1$--$m_2$ bins \cite{Heinzel:2024jlc} (see Appendix~\ref{section: Models} for more details of the model and Appendix~\ref{section: Impact of GW231123} for results including GW231123). The peaks in the $m_2$ distribution are more prominent compared to the parametric models \cite{Farah:2023swu}, suggesting current default models \cite{LIGOScientific:2025pvj} need to be revised. The distributions do not decline sharply at low and high masses, reflecting the agnostic \textsc{PixelPop} prior \cite{Heinzel:2024jlc}, unlike the assumption of strict minima and maxima in the parametric models. Instead, in Figure~\ref{figure: observations} we show the 99\%, 99.9\%, and 99.99\% percentile masses, computed from the marginal cumulative distributions for $m_1$ and $m_2$ from \textsc{PixelPop}, which from GWTC-4 (GWTC-3) are $33_{-3}^{+3}M_\odot$, $49_{-7}^{+16}M_\odot$, and $82_{-23}^{+48}M_\odot$ ($33_{-3}^{+5}M_\odot$, $52_{-12}^{+45}M_\odot$, and $91_{-32}^{+50}M_\odot$) for $m_2$, respectively (see Table~\ref{table: percentiles} later for a summary). Although the 99.9\% $m_2$ is fairly well constrained and consistent with the $m_2$ cutoff in the \textsc{Single Power Law + 2 Peaks + Cutoff} model, the greater flexibility of \textsc{PixelPop} is balanced against larger uncertainties and the presence of mergers with $m_2>50M_\odot$ is not ruled out.

However, the \textsc{PixelPop} $m_2$ distribution declines more steeply than that for $m_1$ and is steeper above $\sim30M_\odot$ than over $\sim10$--20$M_\odot$. The bin-to-bin gradients (logarithmic density over logarithmic mass) for $30M_\odot<m_2<60M_\odot$ (chosen as it has the same logarithmic width as the 10--20$M_\odot$ interval) and $30M_\odot<m_2<150M_\odot$ (chosen as $150M_\odot$ is the highest \textsc{PixelPop} bin edge) are respectively $-6.5_{-2.1}^{+2.0}$ and $-6.3_{-3.5}^{+3.0}$ ($-6.6_{-2.7}^{+2.5}$ and $-6.3_{-4.1}^{+3.6}$) for GWTC-4 (GWTC-3). These are steeper than the slope $-3.1_{-2.1}^{+2.0}$ ($-3.3_{-2.7}^{+2.4}$) over $10M_\odot<m_2<20M_\odot$ at 97\% and 93\% (94\% and 87\%) posterior credibility (see Table~\ref{table: slopes} later for a summary). Therefore, while no strict mass cutoff is inferred with \textsc{PixelPop}, the number of high-mass BHs in binary mergers declines more steeply for secondary than primary components, similar to \textsc{Single Power Law + 2 Peaks + Cutoff} results.

\subsection{Predictive checks}
\label{section: Predictive checks}

Given the differences between the \textsc{Single Power Law + 2 Peaks + Cutoff} and \textsc{PixelPop} results, we test whether the former is a reasonable model to begin with. We use a posterior predictive check (PPC) \cite{Romero-Shaw:2022ctb} but, unlike common approaches in GW population studies, we compare observed and predicted sources self consistently (see Ref.~\cite{Miller:2024sui} for a discussion). For each observed event, we select the parameter estimate with the highest likelihood\footnote{We use this term instead of ``maximum likelihood'' as the sample with the highest likelihood among a set of posterior samples has no guarantee of being the global maximum.} among its posterior samples. We compare the observed highest-likelihood secondary masses to predictions from the \textsc{Single Power Law + 2 Peaks + Cutoff} model marginalized over posterior uncertainty from GWTC-4 by: (1) selecting one population posterior sample, (2) selecting a catalog (equal in size to the real one) of simulated detections from this population, (3) selecting the highest-likelihood sample for each, and (4) repeating many times; more details of the simulated catalogs using full Bayesian parameter estimation (PE) results from Refs.~\cite{Vitale:2025lms, Wolfe:2025yxu} are given in Appendix~\ref{section: Simulations}.

\begin{figure}
\centering
\includegraphics[width=0.9\columnwidth]{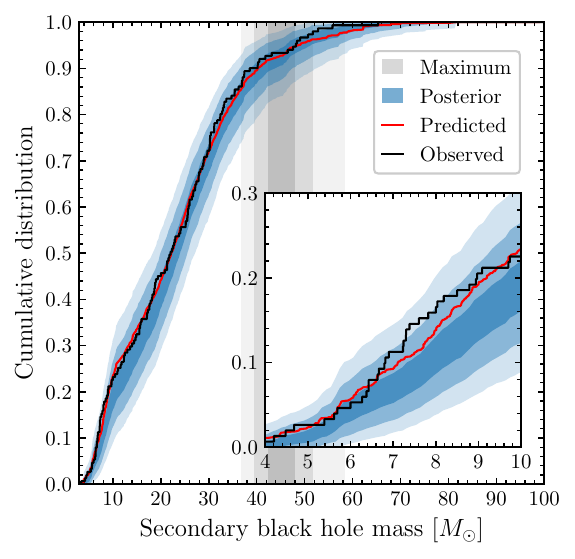}
\caption{Self-consistent data-level PPC showing the cumulative distribution of highest-likelihood secondary BH masses for: events in GWTC-4 (black); central 50\%, 90\%, and 99\% posterior credible regions for simulated observations based on the \textsc{Single Power Law + 2 Peaks + Cutoff} fit to GWTC-4 (darker to lighter blue); and the highest-likelihood population prediction (red). Central 50\%, 90\%, and 99\% posterior credible intervals for the astrophysical maximum secondary mass (darker to lighter gray) are included for reference.}
\label{figure: ppc}
\end{figure}

The results of this check are shown in Figure~\ref{figure: ppc}. We find that the observed events lie within the expected data-level cumulative distribution across (highest-likelihood) secondary BH masses, suggesting that the differences between the parametric and nonparametric results in Figure~\ref{figure: observations} do not lead to significantly discrepant observable properties. In fact, the mass range that sees the largest deviation is well away from a possible PISN gap, lying at $m_2\sim7$--$8M_\odot$, where the observed distribution lies above the 95th percentile of predictions; elsewhere, it lies within at most the central 90\% credible region. The observed distribution also turns over around $\sim$40--50$M_\odot$, where the $m_2$ cutoff is inferred to lie. We also perform a more stringent one-sample Kolmogorov--Smirnov (KS) test by comparing the distribution of observed highest-likelihood secondary masses to only the predicted distribution with the highest population likelihood among the set of population posterior samples. Under the null hypothesis, the observations should fall uniformly against the cumulative predicted distribution, for which the KS $p$-value is 0.97.

\begin{figure}
\centering
\includegraphics[width=0.9\columnwidth]{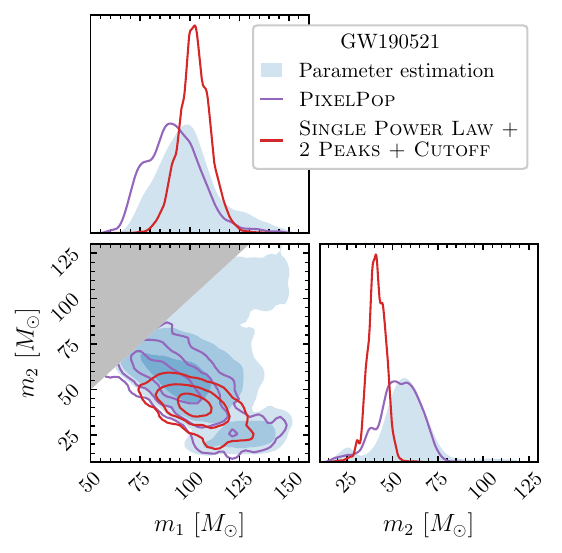}
\caption{Posterior distributions for the source masses of GW190521 from the original LVK PE (blue) and population-informed measurements using the \textsc{PixelPop} (purple) and \textsc{Single Power Law + 2 Peaks + Cutoff} (red) models. The lower-left panel shows the 50\%, 90\%, and 99\% credible regions for the joint posteriors of primary ($m_1$) and secondary ($m_2$) masses, with the gray region excluded by the definition $m_1 \geq m_2$, and the diagonal panels show the individual marginal posteriors.}
\label{figure: gw190521}
\end{figure}

As a final check, we consider the event with the most massive highest-likelihood secondary BH---GW190521 \cite{LIGOScientific:2020iuh, LIGOScientific:2020ufj} (recall GW231123 is excluded from our analyses). In Figure~\ref{figure: gw190521}, we compare the source masses from the original LVK PE results to our population-informed measurements \cite{T1900895, T2100301, Moore:2021xhn} using the \textsc{Single Power Law + 2 Peaks + Cutoff} and \textsc{PixelPop} models. The \textsc{PixelPop} posterior contains much more support for $m_2>50M_\odot$, consistent with the lack of a sharp truncation in Figure~\ref{figure: observations}. This implies that even if the secondary BH of GW190521 lies below a PISN mass gap, it is not apparent without stronger model assumptions.

\section{Validating GWTC-4 constraints}
\label{section: Validating GWTC-4 constraints}

\begin{figure*}
\centering
\includegraphics[width=1\textwidth]{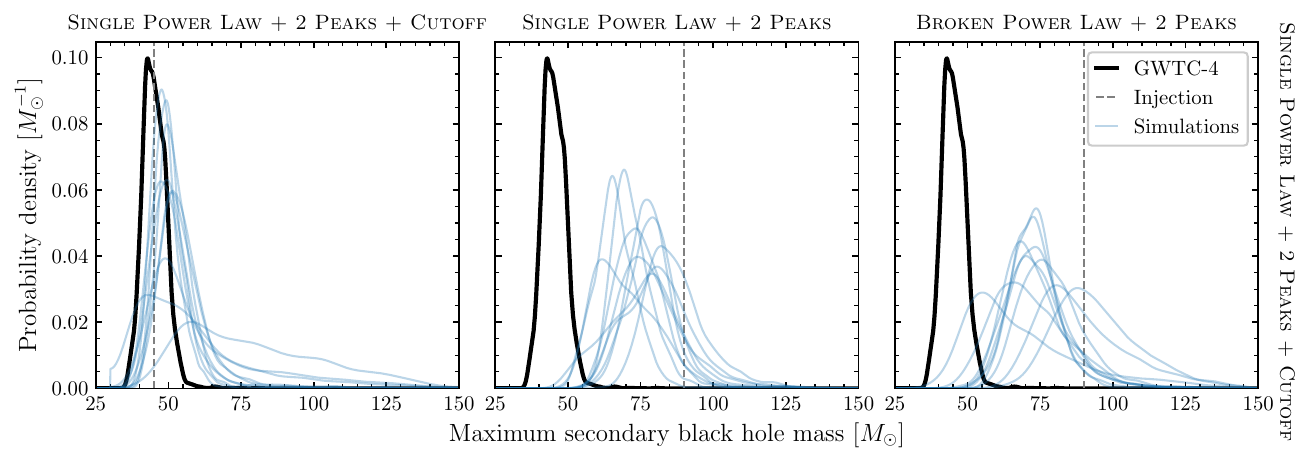}
\includegraphics[width=1\textwidth]{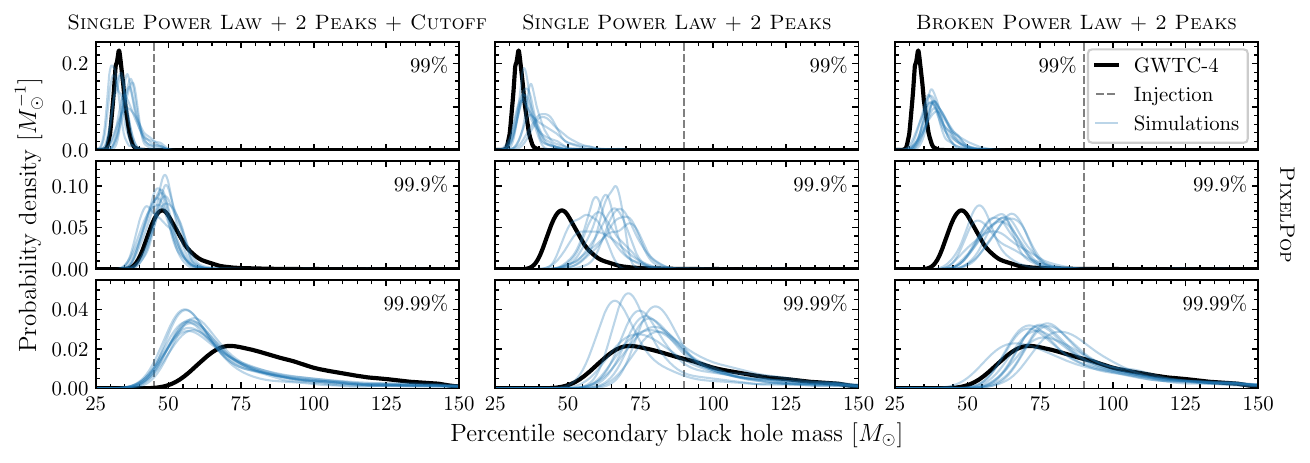}
\caption{Constraints on the mass cutoff for secondary BHs ($m_2$) inferred from GWTC-4 without GW231123 (black) and from simulated catalogs containing 150 O4-like GW events (blue). The true simulated values for the maximum $m_2$ are marked with dashed gray lines: $45M_\odot$ in the left panel and $90M_\odot$ in the middle and right panels. Each column represents a different simulated population indicated in the column titles, with 10 catalog realizations for each. The top row shows posteriors on the maximum $m_2$ when the real and simulated catalogs are analyzed with the parametric \textsc{Single Power Law + 2 Peak + Cutoff} population model. The bottom rows show the posterior 99\%, 99.9\%, and 99.99\% population-level $m_2$ percentiles inferred from each catalog with the nonparametric \textsc{PixelPop} model.}
\label{figure: mgap}
\end{figure*}

We now test whether or not: (a) the GWTC-4 constraints on the mass cutoff are as expected, including the strength of model preference, when such a cutoff is present in the underlying population; (b) a mass cutoff can be spuriously inferred from the catalog when there is no such feature; (c) a steepening decline of the primary BH mass ($m_1$) spectrum can be more easily confused for a secondary-mass ($m_2$) truncation; and (d) populations with a cutoff can be identified and distinguished from those without using \textsc{PixelPop}.

To do so, we simulate GW catalogs of 150 events with full Bayesian PE from Refs.~\cite{Vitale:2025lms, Wolfe:2025yxu}, commensurate with the GWTC-4 catalog. We consider three different underlying populations according to the three parametric models in Table~\ref{table: models}, with parameters selected consistent with results from GWTC-4 when analyzed with each of those models (see Appendix~\ref{section: Simulations}). We repeat each analysis on 10 different catalog realizations to assess variations in results. Full details on an efficient bootstrap procedure we use to generate these catalogs by reusing previous PE results for simulated sources are given in Appendix~\ref{section: Simulations}.

\subsection{Do GWTC-4 results match expectations?}
\label{section: Do GWTC-4 results match expectations?}

We start with catalogs simulated according to a \textsc{Single Power law + 2 Peaks + Cutoff} population with a mass cutoff at $45M_\odot$. The posteriors on the $m_2$ cutoff for the 10 catalogs when analyzed with that same model are shown in the left-hand panel of the top row in Figure~\ref{figure: mgap}. We find that the posteriors are always consistent with the injected gap location and indeed cluster around the posterior inferred from the catalog of real data, with medians ranging from $\approx48$--$53M_\odot$ for all but one of our simulations that has median $\approx75M_\odot$. In all cases the posterior median overestimates the true value, which may suggest that the true cutoff is lower than $\approx45M_\odot$ implied by inference from GWTC-4. The widths of the 90\% posterior CIs are typically $\approx15$--$30M_\odot$, but for three out of the 10 catalogs are $>60M_\odot$. Compared to a CI of $12M_\odot$ from the analysis of GWTC-4, this implies that the real measurement is consistent with only the best-constrained results among the small number of catalog realizations we consider. However, the simulations also demonstrate that a well-determined mass cutoff is not a guaranteed realization, which only three out of our 10 catalogs constrain to within $\pm10M_\odot$.

We also analyze these same 10 catalogs with the \textsc{Single Power Law + 2 Peaks} and \textsc{Broken Power Law + 2 Peaks} models---neither of which allow for independent $m_2$ and $m_1$ maxima---to test whether a Bayesian model comparison correctly favors the \textsc{Single Power law + 2 Peaks + Cutoff} model from which the catalogs were simulated. None of our simulations produce Bayes factors close to the GWTC-4 result, with the correct model favored at most by $\log_{10}\mathcal{B}\approx0.8$; cf. Table~\ref{table: models}. Four out of 10 catalogs have no preference within the evidence uncertainties reported from nested sampling ($\pm0.1$) and for one catalog the \textsc{Broken Power Law + 2 Peaks} model is actually preferred over the true model, but only by $\log_{10}\mathcal{B}\approx0.3$. The weaker model preference compared to GWTC-4 results is in line with broader posteriors on the $m_2$ cutoff from our simulations.

\subsection{Can a cutoff be spuriously inferred?}
\label{section: Could a cutoff be spuriously inferred?}

We simulate another 10 catalogs, this time with sources drawn from a \textsc{Single Power Law + 2 Peaks} population. In particular, both $m_1$ and $m_2$ maxima are $90M_\odot$ (chosen based on the GWTC-4 constraints in Figure~\ref{figure: observations}). The $m_2$ cutoff posteriors inferred when these catalogs are instead analyzed with the \textsc{Single Power law + 2 Peaks + Cutoff} model allowing for independent $m_1$ and $m_2$ maxima are shown in the middle panel in the top row of Figure~\ref{figure: mgap}. None of these measurements are qualitatively compatible with the analysis of GWTC-4, implying that we are unlikely to spuriously infer a PISN cutoff as low as $\sim45M_\odot$ when it does not exist. The lowest inferred posterior median is $\approx66M_\odot$. Note that the \textit{marginal} one-dimensional posteriors for the maximum $m_2$ are systematically lower than the injected value because the prior on the $m_2$ maximum ensures it is always lower than that of the $m_1$ maximum, but the injected value is always consistent with the \textit{joint} posterior on both maxima. In most cases, the \textsc{Single Power Law + 2 Peaks} is correctly preferred over the \textsc{Single Power law + 2 Peaks + Cutoff} model, though again not substantially: seven out of 10 mock catalogs favor it with $0.2\lesssim\log_{10}\mathcal{B}\lesssim0.6$, the other three cases being undecided within the estimated evidence error ($\pm0.1$).

\subsection{Can a declining density appear as a cutoff?}
\label{section: Can a declining density appear as a cutoff?}

We also test whether a more steeply declining $m_1$ power law above a break mass $m_1>34M_\odot$ (this mass chosen to be consistent with our GWTC-4 analyses; see Appendix~\ref{section: Simulations}) is more likely to lead to spurious measurement of an $m_2$ cutoff, simulating 10 catalogs from a \textsc{Broken Power Law + 2 Peaks} population. The posteriors for the $m_2$ maximum inferred when using the incorrect \textsc{Single Power law + 2 Peaks + Cutoff} model are shown in the right-hand panel in the top row of Figure~\ref{figure: mgap}. These results are similar to the \textsc{Single Power Law + 2 Peaks} simulations, except several are broader and skewed towards lower masses, as expected. From the catalog producing the posterior with the lowest median, we measure the $m_2$ maximum to be $63_{-16}^{+48}M_\odot$, which overlaps more with the GWTC-4 result than the results from the other \textsc{Broken Power Law + 2 Peaks} catalogs and those from the \textsc{Single Power Law + 2 Peaks} catalogs. On the other hand, from the \textsc{Single Power law + 2 Peaks + Cutoff} simulations, there is one catalogs that yields a similar $m_2$ cutoff measurement of $75_{-25}^{+50}M_\odot$, suggesting that an $m_2$ cutoff could have been more easily confused with a declining mass spectrum had GWTC-4 been realized differently. But for the actual realization, the power-law slope of the underlying population at high masses would have to be even steeper than simulated ($<-4$) for the two scenarios to be less distinguishable than in Figure~\ref{figure: mgap}. That said, four of these catalogs do not show model preference within the estimated error, so while the misspecification we introduce does not lead to incorrectly favoring an $m_2$ truncation, the catalogs also do not always confidently favor the correct model either.

\subsection{Are nonparametric constraints as expected?}
\label{section: Are nonparametric constraints as expected?}

\begin{table*}
\caption{Population percentiles for secondary BH masses $m_2/M_\odot$ inferred using \textsc{PixelPop} from the LVK catalogs (top two rows) and simulated catalogs of 150 (middle three rows) or 300 (bottom three rows) O4-like events. Values are posterior medians and in brackets are widths of 90\% CIs. For simulations, ranges cover 10 catalog realizations.}
\centering
\setlength{\tabcolsep}{5pt}
\begin{tabular}{cccc}
\hline\hline
Catalog &
99\% &
99.9\% &
99.99\% \\
\hline\hline
GWTC-3 &
33 (8) &
52 (57) &
91 (82) \\
GWTC-4 &
33 (6) &
49 (23) &
82 (71) \\
\hline
\textsc{Single Power Law + 2 Peaks + Cutoff} &
31--37 (7--14) &
45--50 (12--16) &
58--64 (54--76) \\
\textsc{Single Power Law + 2 Peaks} &
35--43 (8--20) &
56--68 (13--25) &
69--88 (52--67) \\
\textsc{Broken Power Law + 2 Peaks} &
37--40 (10--18) &
55--65 (20--28) &
76--87 (59--73) \\
\hline
\textsc{Single Power Law + 2 Peaks + Cutoff} &
33--37 (4--7) &
43--47 (9--13) &
56--61 (37--60) \\
\textsc{Single Power Law + 2 Peaks} &
33--43 (8--12) &
60--75 (12--19) &
73--84 (31--52) \\
\textsc{Broken Power Law + 2 Peaks} &
36--41 (9--11) &
58--68 (12--24) &
70--84 (44--54) \\
\hline\hline
\end{tabular}
\label{table: percentiles}
\end{table*}

\begin{table*}
\caption{Slopes (logarithmic density over logarithmic mass) of secondary-mass ($m_2$) distributions between 10--20$M_\odot$ (second column), 30--60$M_\odot$ (third column), and 30--150$M_\odot$ (fourth column) inferred by \textsc{PixelPop} from the LVK catalogs (top two rows) and simulated catalogs of 150 (middle three rows) or 300 (bottom three rows) O4-like events. Values are posterior medians, in brackets are widths of 90\% CIs, and percentages are posterior probabilities that slopes are steeper than over 10--20$M_\odot$. For simulated catalogs, ranges cover 10 catalog realizations.}
\centering
\setlength{\tabcolsep}{5pt}
\begin{tabular}{cccc}
\hline\hline
Catalog &
10--20$M_\odot$ &
30--60$M_\odot$ &
30--150$M_\odot$ \\
\hline\hline
GWTC-3 &
3.3 (5.1) &
6.6 (5.2) : 94\% &
6.3 (7.7) : 87\% \\
GWTC-4 &
3.1 (4.2) &
6.5 (4.1) : 97\% &
6.3 (6.5) : 93\% \\
\hline
\textsc{Single Power Law + 2 Peaks + Cutoff} &
1.8--4.2 (4.9--6.4) &
6.9--8.9 (7.2--8.7) : 92--97\% &
7.8--8.8 (8.5--11.2) : 91--98\% \\
\textsc{Single Power Law + 2 Peaks} &
1.4--3.9 (4.8--6.4) &
3.1--5.7 (5.0--6.9) : 58--91\% &
6.8--7.8 (7.7--9.8) : 89--98\% \\
\textsc{Broken Power Law + 2 Peaks} &
0.3--3.2 (4.5--6.6) &
4.3--6.2 (5.2--8.0) : 67--98\% &
6.6--8.1 (7.6--9.9) : 92--99\% \\
\hline
\textsc{Single Power Law + 2 Peaks + Cutoff} &
2.6--3.8 (4.1--5.3) &
7.7--8.8 (6.3--7.4) : 95--98\% &
7.8--8.5 (8.2--8.7) : 93--98\% \\
\textsc{Single Power Law + 2 Peaks} &
2.1--3.9 (4.3--5.5) &
3.4--5.2 (4.6--6.0) : 60--87\% &
7.0--7.9 (7.2--8.7) : 89--98\% \\
\textsc{Broken Power Law + 2 Peaks} &
1.0--3.8 (4.3--5.5) &
4.3--6.4 (4.8--5.8) : 60--94\% &
7.0--8.1 (7.8--8.7) : 92--99\% \\
\hline\hline
\end{tabular}
\label{table: slopes}
\end{table*}

Next, we reanalyze all 30 catalogs above but with the nonparametric \textsc{PixelPop} population model. In the bottoms rows of Figure~\ref{figure: mgap}, we show posteriors for the same $m_2$ population percentiles as in Figure~\ref{figure: observations} (see also Figure~\ref{figure: pixelpop predictions} later for the population distributions). We find that the percentiles inferred from the \textsc{Single Power Law + 2 Peaks + Cutoff} catalogs are most similar to those inferred from GWTC-4, particularly 99\% and 99.9\%, as summarized in Table~\ref{table: percentiles}. The 99.9\% $m_2$ values most closely recover the injected $m_2$ maximum of $45M_\odot$, with medians 45--50$M_\odot$ similar to GWTC-4 ($49M_\odot$) but smaller 90\% CIs 12--16$M_\odot$ than GWTC-4 ($23M_\odot$); for the analyses of the \textsc{Single Power law + 2 Peaks} and \textsc{Broken Power Law + 2 Peaks} catalogs, the respective medians---56--68$M_\odot$ and 55--65$M_\odot$---differ from the GWTC-4 result, but have similar uncertainties---13--25$M_\odot$ and 20--28$M_\odot$. The constraints at the 99.99\% percentile from these latter two populations are more similar to the GWTC-4 result, while those for the \textsc{Single Power Law + 2 Peaks + Cutoff} catalogs are dissimilar and show the least variation, suggesting the constraints on the very tail of this $m_2$ distribution depend more on prior choices due to the lack of detections above the $45M_\odot$ cutoff.

\begin{figure*}
\centering
\includegraphics[width=1\textwidth]{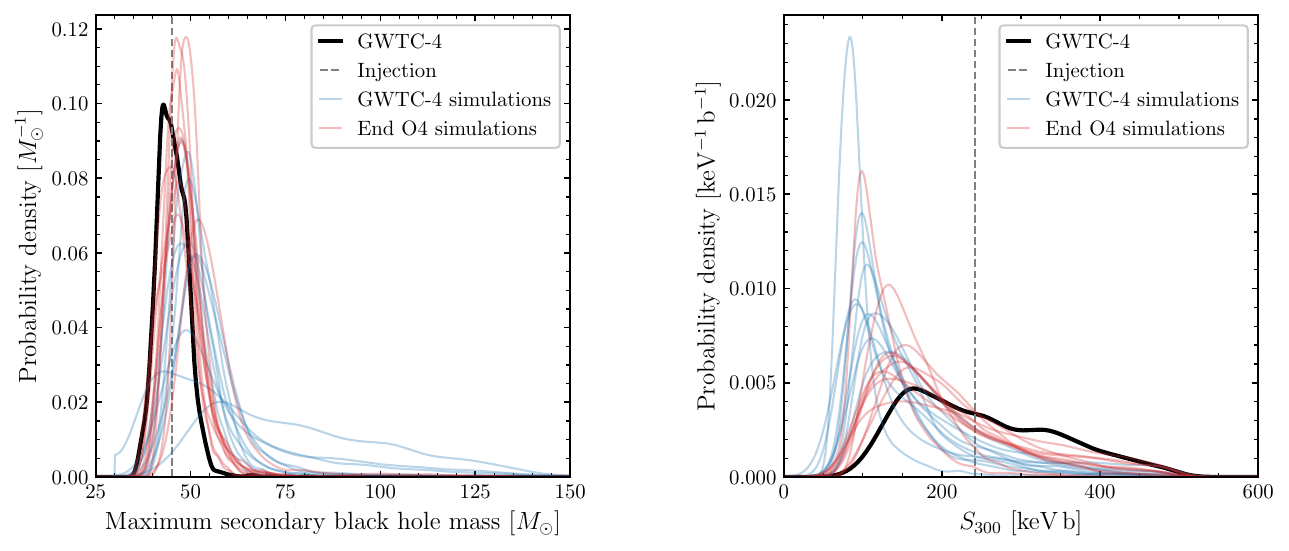}
\caption{Posterior distributions for the maximum secondary BH mass (left panel) and the S-factor for the $^{12}\mathrm{C}(\alpha,\gamma)^{16}\mathrm{O}$ reaction at a temperate 300\,keV (right panel), assuming this maximum mass is the lower edge of the PISN mass gap, inferred using the \textsc{Single Power Law + 2 Peaks + Cutoff} model. Results from GWTC-4 (solid black) are used to select a simulated population for this model (dashed gray), from which 10 catalogs are analyzed, each of 150 (blue) and 300 (red) O4-like events.}
\label{figure: pisn-predictions}
\end{figure*}

As in Section~\ref{section: Nonparametric results}, we compare the slope of the merger rate over higher and lower $m_2$ ranges, with results summarized in Table~\ref{table: slopes}. In general, the uncertainties from our simulated catalogs are larger than from GWTC-4. The posterior medians of the slopes for $10<m_2/M_\odot<20$ are consistent with the GWTC-4 result for all three simulated populations, though the \textsc{Broken Power Law + 2 Peaks} model is simulated with a shallower slope in this mass range than the other models, which is correctly recovered by \textsc{PixelPop}. Over $30<m_2/M_\odot<60$, the median slopes inferred from the \textsc{Single Power Law + 2 Peaks + Cutoff} catalogs are typically steeper than from GWTC-4, but those from the other catalogs are typically shallower than from GWTC-4. Over the broader range $30<m_2/M_\odot<150$, the GWTC-4 median is typically shallower than results from catalogs of all three simulated populations. From the \textsc{Single Power Law + 2 Peaks + Cutoff} catalogs, \textsc{PixelPop} identifies that the $m_2$ slope over 30--60$M_\odot$ is steeper than over 10--20$M_\odot$ with posterior probabilities ($\gtrsim92\%$) consistently more similar to the GWTC-4 result (97\%) than the other simulated catalogs ($\gtrsim60\%$); on the other hand, \textsc{PixelPop} reports posterior probabilities for a steeper slope over 30--150$M_\odot$ than 10--20$M_\odot$ compatible with the GWTC-4 result for all simulated catalogs.

Overall, these results suggest that models without an $m_2$ cutoff may be missing a declining density of sources between 30--60$M_\odot$, but that a strict cutoff may lead to too steep a decline, in agreement with our obsevations in Sections~\ref{section: Nonparametric results} and \ref{section: Predictive checks}. In Appendix~\ref{section: Impact of single-event uncertainty}, we assess the impact of single-event measurement uncertainty on identifying the $m_2$ cutoff in catalogs from the \textsc{Single Power Law + 2 Peaks + Cutoff} population with \textsc{PixelPop}.

\section{Projections for the end of O4}
\label{section: Projections for the end of O4}

Moving from current to future GW observations, we project constraints that may be possible by the end of the LVK O4 observing run. We use the same 10 catalog realizations as in Section~\ref{section: Validating GWTC-4 constraints} for each simulated population but add a further 150 events to each, roughly compatible with the total number of confident binary BH mergers that may be detected by the end of O4 \cite{gracedb}.

\subsection{Populations with a pair-instability cutoff}
\label{section: Pair-instability cutoff}

In Figure~\ref{figure: pisn-predictions}, we show the posteriors on the maximum secondary mass when catalogs drawn from the \textsc{Single Power law + 2 Peaks + Cutoff} population are analyzed using that same model. In contrast to the catalogs of 150 events, by the time the catalog has again doubled in size there are no longer such ambiguously broad posteriors. Among the 10 catalog realizations, the posterior medians range between 46--54$M_\odot$ (cf. the simulated value of $45M_\odot$), with 90\% CIs 10--20$M_\odot$, reflecting reductions in uncertainty up to $\lesssim85\%$ and $\gtrsim20\%$.

When these catalogs are instead analyzed with the other two models in Table~\ref{table: models}, we find the correct model is always favored, in both cases now with only one out of the 10 catalogs having preference $\log_{10}\mathcal{B}\approx0.2$ within the estimated error ($\pm0.2$). Compared to the \textsc{Single Power Law + 2 Peaks} (\textsc{Broken Power Law + 2 Peaks}) model, five (four) catalogs correctly prefer the \textsc{Single Power Law + 2 Peaks + Cutoff} model with $\log_{10}\mathcal{B}\gtrsim1$, but the largest Bayes factor $\log_{10}\mathcal{B}\approx1.5$ out of all these analyses is still much less than in our analyses of GWTC-4.

We also show the derived posteriors on the S-factor $S_{300}$ in Figure~\ref{figure: pisn-predictions}. Note that for the larger catalogs, the posterior is counterintuitively broadened, but at the same time becomes more consistent with the simulated value. This is because the edges of the mass gap are not altered much at higher $^{12}\mathrm{C}(\alpha,\gamma)^{16}\mathrm{O}$ reaction rates; cf. Figure~5 of Ref.~\cite{Farmer:2020xne}. Therefore, if the pair-instability cutoff indeed falls at $\sim45M_\odot$, future GW catalogs will likely only be able to constrain a lower bound. Our 10 realizations of O4-like catalogs containing 150 events span $S_{300}>69$--104\,(68--90)\,keV\,b at 90\% (99\%) credibility, compared to $S_{300}>89$--126\,(71--98)\,keV\,b for our catalogs of 300 events.

\subsection{Populations without a pair-instability cutoff}
\label{section: Populations without a pair-instability cutoff}

Although the results of Section~\ref{section: Validating GWTC-4 constraints} suggest that the  measured $m_2$ cutoff is unlikely to be a spurious false measurement, we next consider the catalogs simulated from the \textsc{Single Power Law + 2 Peaks} population, for which both BH maxima are 90$M_\odot$, and analyze them with (\textsc{Single Power law + 2 Peaks + Cutoff}) and without (\textsc{Single Power Law + 2 Peaks}) an independent $m_2$ truncation. For most catalogs, the latter model is correctly preferred according to the Bayes factor, with largest preference $\log_{10}\mathcal{B}\approx0.8$, except in two cases for which it is within the estimated error ($\pm0.2$).

\begin{figure}
\centering
\includegraphics[width=1\columnwidth]{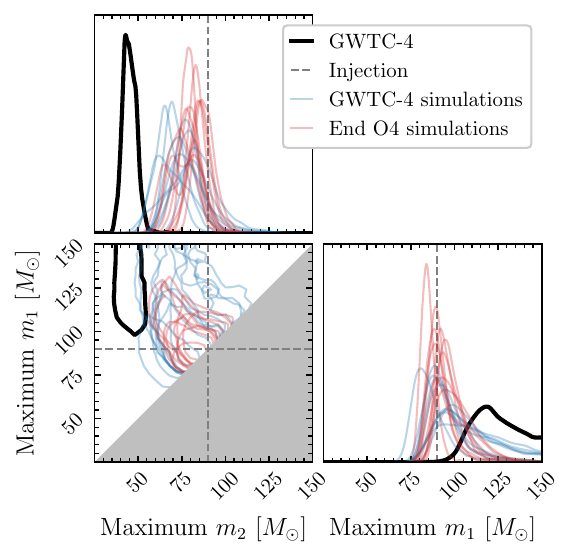}
\caption{Posterior distributions for the maximum primary ($m_1$) and secondary ($m_2$) BH masses inferred by the \textsc{Single Power Law + 2 Peaks + Cutoff} model from GWTC-4 (solid black) and from 10 simulated catalogs, each with 150 (blue) and 300 (red) O4-like events from the \textsc{Single Power Law + 2 Peaks} population that has a common maximum mass of 90$M_\odot$ (dashed gray). The bottom-left panel shows the 90\% posterior credible regions for the joint posteriors, with the gray region excluded by the constraint $m_1 \geq m_2$. The top-left and bottom-right panels show the individual marginal posteriors.}
\label{figure: mmax-predictions}
\end{figure}

In Figure~\ref{figure: mmax-predictions}, we show the results of the former analysis. In particular, we find that as the maximum primary and secondary BH mass are equal in the true population, the posteriors indeed tend to cluster at equal maxima (the diagonal in the bottom-left panel of Figure~\ref{figure: mmax-predictions}). This is in contrast to the clearly distinct constraints from the GWTC-4 analysis. For all catalogs, we find posterior medians $\gtrsim75M_\odot$ for the maximum secondary mass and 90\% CIs up to $\sim60\%$ narrower than the catalogs of 150 events, but one that is $\sim25\%$ broader. Nonetheless, comparing the inferred maximum $m_2$ between Figures~\ref{figure: pisn-predictions} and \ref{figure: mmax-predictions} confirms that a continued measurement at $\sim45M_\odot$ is increasingly unlikely to be spurious.

\begin{figure}
\centering
\includegraphics[width=1\columnwidth]{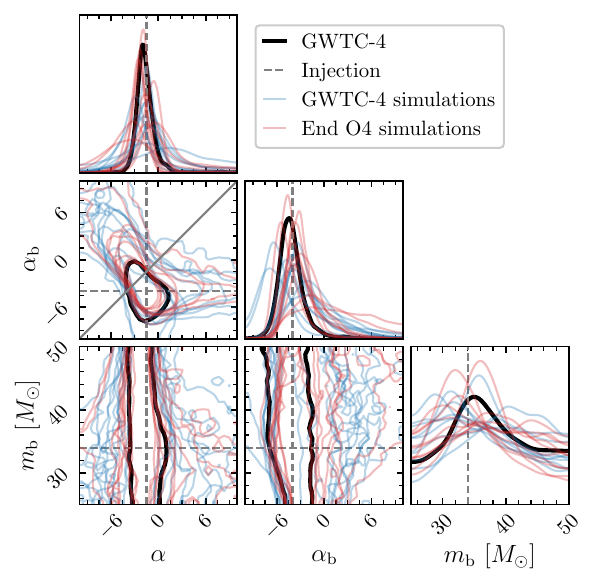}
\caption{Posterior distributions for parameters of the \textsc{Broken Power Law + 2 Peaks} model: the location $m_\mathrm{b}$ of the break in the primary-mass power law, the power-law slope $\alpha_\mathrm{b}$ above the break, and the slope $\alpha$ below the break. Results from GWTC-4 (black) are used to select a simulated population (dashed gray) from which 10 catalogs with 150 (blue) and 300 (red) O4-like events are drawn and analyzed with that same model. Panels on the diagonal show marginal posteriors and other panels show 90\% credible regions for two-dimensional posteriors. Equal power-law slopes $\alpha=\alpha_\mathrm{b}$ (solid gray) are marked for reference in the middle-left panel.}
\label{figure: break-predictions}
\end{figure}

We also consider the catalogs drawn from the \textsc{Broken Power Law + 2 Peaks} population to test how well we might measure the properties of the mass distribution if the true population were closer to this model in reality. Out of the 10 catalogs of 300 events, six favor the correct model with $\log_{10}\mathcal{B}\lesssim1$, the remaining cases being undecided within the estimated error ($\pm0.2$). In Figure~\ref{figure: break-predictions}, we show the posteriors for the location $m_\mathrm{b}$ of the break in the $m_1$ power law, the power-law slope $\alpha_\mathrm{b}$ above the break (higher $m_1$), and the slope $\alpha$ below the break (lower $m_1$) from analyses of GWTC-4 and simulated catalogs of 150 and 300 events. The GWTC-4 results are typically better constrained than our simulations of similar catalogs, especially the power-law slopes, which may suggest those constraints are driven by features in the data not accounted for by the model. Even when the model is correctly specified, we are unable to confidently constrain a broken power law in catalogs of 300 events, as indicated by the broad $m_\mathrm{b}$ posteriors (bottom-right panel of Figure~\ref{figure: break-predictions})---e.g., the 90\% posterior CIs have widths $\sim20M_\odot$, compared to the prior width of $25M_\odot$---and degeneracy between $\alpha$ and $\alpha_\mathrm{b}$ (middle-left panel of Figure~\ref{figure: break-predictions})---e.g., we only find $\alpha>\alpha_\mathrm{b}$ at posterior credibility 28--94\%.

\subsection{Cosmology projections}
\label{section: Cosmology projections}

\begin{figure*}
\centering
\includegraphics[width=1\textwidth]{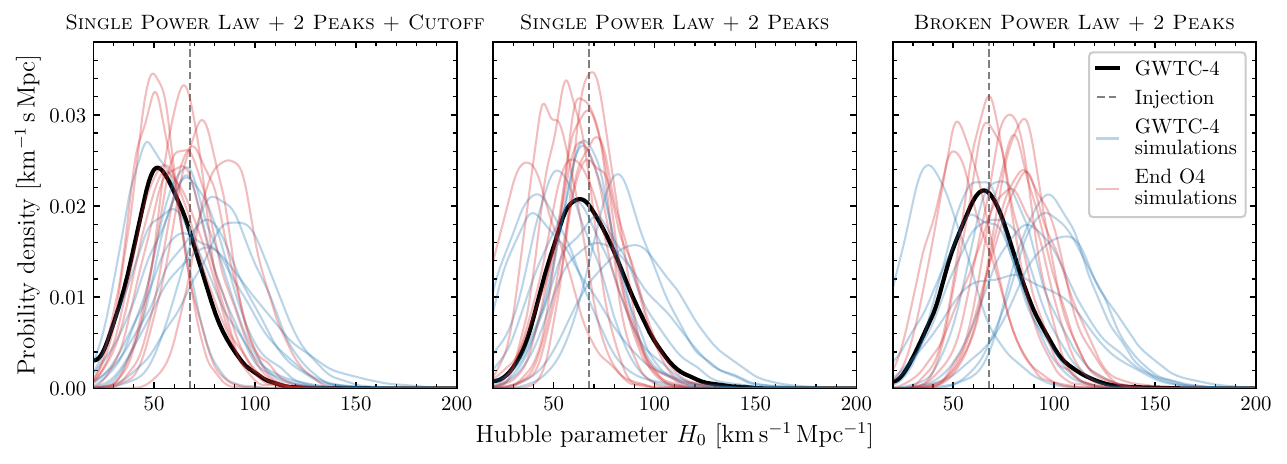}
\caption{Posterior distributions for the Hubble parameter $H_0$ inferred using the GW-only spectral-siren method with the \textsc{Single Power Law + 2 Peaks + Cutoff} (left), \textsc{Single Power Law + 2 Peaks} (middle), and \textsc{Broken Power Law + 2 Peaks} (right) parametric population models. All simulated values are $H_0=67.74\,\mathrm{km}\,\mathrm{s}^{-1}\,\mathrm{Mpc}^{-1}$ \cite{Planck:2015fie} (dashed gray). Other elements are as in previous figures.}
\label{figure: h0-predictions}
\end{figure*}

For each of the three populations above, we analyze all simulated catalogs again but now also inferring the Hubble parameter $H_0$ as in Section~\ref{section: Parametric results}. In each case, the true population has $H_0=67.74\,\mathrm{km}\,\mathrm{s}^{-1}\,\mathrm{Mpc}^{-1}$ \cite{Planck:2015fie} and the model used for inference matches the simulated population (except its parameters are inferred). The results are shown in Figure~\ref{figure: h0-predictions}.

The posteriors from our simulated catalogs of 150 events are comparable to the GWTC-4 posteriors, with the widths of the 90\% posterior CIs among each of the 10 catalog realizations ranging between 45--89, 51--82, and 56--95\,km\,s$^{-1}$\,Mpc$^{-1}$ for the \textsc{Single Power law + 2 Peaks + Cutoff}, \textsc{Single Power Law + 2 Peaks}, and \textsc{Broken Power Law + 2 Peaks} populations, respectively; cf. 56, 63, and 63\,km\,s$^{-1}$\,Mpc$^{-1}$ for the analyses of GWTC-4.

These uncertainties are reduced and span 38--54, 37--53, and 44--57\,km\,s$^{-1}$\,Mpc$^{-1}$, respectively, from analyses of our simulated catalogs of 300 events, suggesting a reduction of up to $\sim50\%$ in the $H_0$ uncertainty from spectral-siren analyses alone by the end of O4. The presence of an $m_2$ cutoff in the \textsc{Single Power law + 2 Peaks + Cutoff} catalogs does not necessarily result in better $H_0$ constraints than from the \textsc{Single Power Law + 2 Peaks} catalogs without one, but the best $H_0$ constraints from both these populations are typically better than from \textsc{Broken Power Law + 2 Peaks} catalogs. However, the relative uncertainty (ratio of posterior 90\% CI to median) in $H_0$ can still be as large as $\mathcal{O}(100\%)$.

\subsection{Nonparametric projections}
\label{section: Nonparametric projections}

\begin{figure*}
\centering
\includegraphics[width=1\textwidth]{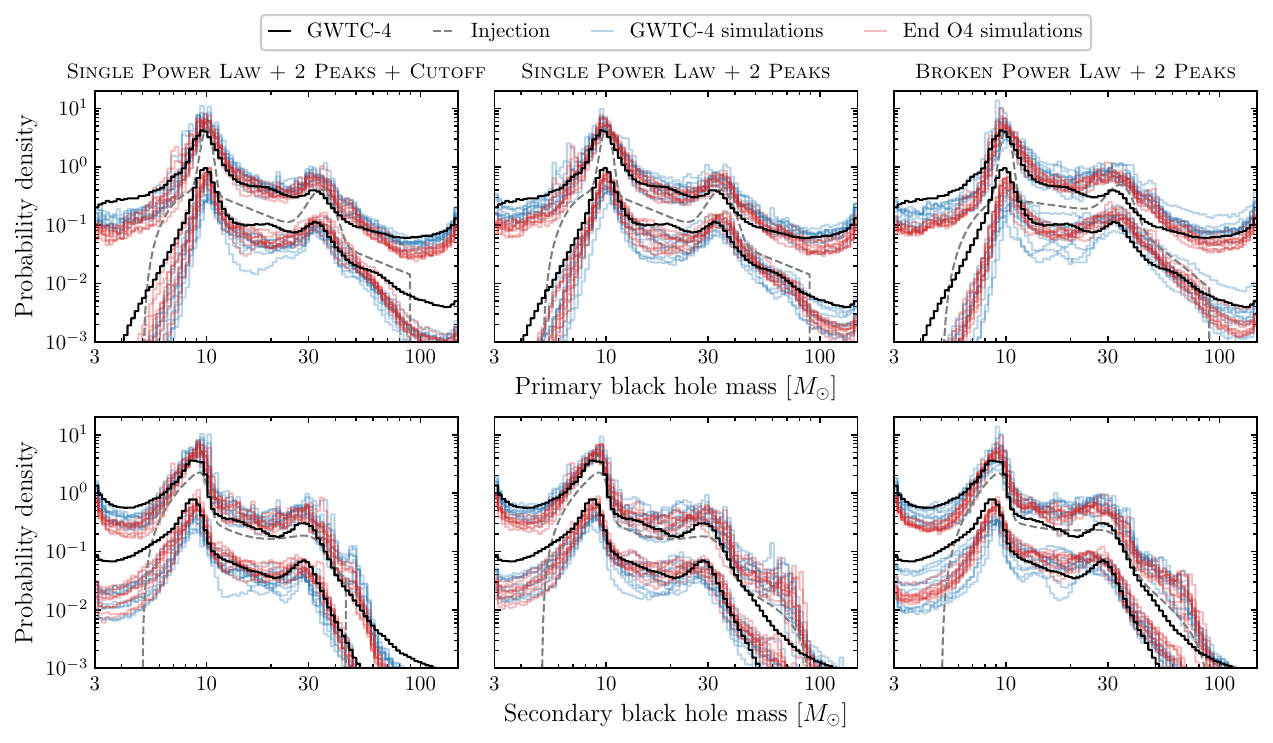}
\vspace{-20pt}
\caption{Population distributions of primary (top row) and secondary (bottom row) BH masses inferred by \textsc{PixelPop} in terms of 90\% CIs for probability densities of logarithmic mass. Each column shows results for analyses of 10 catalogs drawn from simulated populations (dashed gray), based on fits to GWTC-4, each containing 150 (blue) or 300 (red) O4-like events. The \textsc{PixelPop} results for GWTC-4 (solid black) are included for reference.}
\label{figure: pixelpop predictions}
\end{figure*}

\begin{figure*}
\centering
\includegraphics[width=1\textwidth]{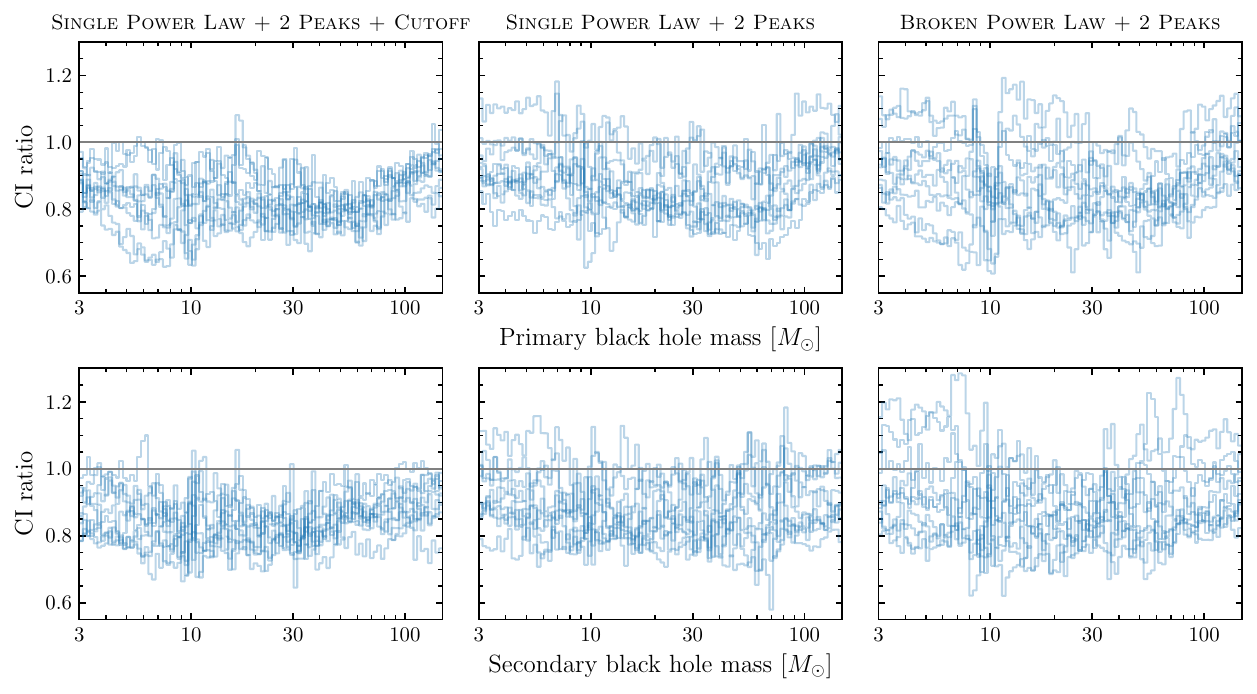}
\vspace{-20pt}
\caption{Ratios of 90\% CIs for binary BH merger rates in primary (top row) and secondary (bottom row) mass bins between simulated catalogs of 150 and 300 events across 10 catalog realizations (blue), as inferred by \textsc{PixelPop}. Ratios below unity (gray) indicate reduced uncertainty with increased catalog size.}
\label{figure: pixelpop ratios}
\end{figure*}

Finally, we repeat the analyses of the three populations with the nonparametric \textsc{PixelPop} model (and fixed cosmological model). In Figure~\ref{figure: pixelpop predictions}, we show the population distributions of primary and secondary BH masses inferred from the simulated catalogs of 150 and 300 events, and compare also to the GWTC-4 result. As in Section~\ref{section: Are nonparametric constraints as expected?}, the predicted \textsc{PixelPop} uncertainties are typically larger than found from GWTC-4, though the overall shape of the population is compatible with our simulations. The \textsc{Single Power Law + 2 Peaks} and \textsc{Broken Power Law + 2 Peaks} catalogs without an independent secondary-mass cutoff tend to overpredict the number of mergers with secondary masses above 40--50$M_\odot$ compared to GWTC-4. On the other hand, the $m_2$ cutoff in the \textsc{Single Power Law + 2 Peaks + Cutoff} population can cause a local overdensity at $m_2\sim45M_\odot$ in the inferred distribution that is not seen in the GWTC-4 result, and leads to an underestimate of the number of mergers with secondary masses above 60--70$M_\odot$ compared to GWTC-4. None of our populations are simulated with a significant peak at $\sim30M_\odot$ in the $m_2$ distribution, but such peaks can nonetheless be present in the inferred populations.

The impact of doubling the catalog size on the inferred merger rate across BH masses is quantified in Figure~\ref{figure: pixelpop ratios}. The widths of the 90\% CIs for merger rates in each primary and secondary mass bin typically reduce by $\sim20\%$ and up to $\lesssim40\%$, but in some instances increase by $\lesssim20\%$. Catalogs of 300 events simulated from the \textsc{Single Power Law + 2 Peaks + Cutoff} population typically produce more consistent constraints and improvements thereof over catalogs of 150 events than when simulated from the \textsc{Single Power Law + 2 Peaks} and \textsc{Broken Power Law + 2 Peaks} populations (but this may also be due to our simulation and resampling procedure described in Appendix~\ref{section: Simulations}).

Our projected constraints on the upper $m_2$ percentiles are again summarized in Table~\ref{table: percentiles}. From the \textsc{Single Power Law + 2 Peaks + Cutoff} catalogs, we find 99.9\% $m_2$ posterior medians ranging between 43--47$M_\odot$ with 90\% CI widths 9--13$M_\odot$, reflecting uncertainties reduced by $\lesssim30\%$ and again demonstrating that this percentile tracks well the simulated $m_2$ cutoff when present. For the \textsc{Single Power Law + 2 Peaks} (\textsc{Broken Power Law + 2 Peaks}) model, the median 99.9\% $m_2$ values span 60--75$M_\odot$ (58--68$M_\odot$) with 90\% CI widths 12--19$M_\odot$ (12--24$M_\odot$), which are again clearly distinguished from the catalogs with an independent $m_2$ cutoff. Projections for the merger-rate slopes over $m_2$ result in similar conclusions as in Section~\ref{section: Are nonparametric constraints as expected?} and are included in Table~\ref{table: slopes} for completeness.

\section{Conclusions}
\label{section: Conclusions}

The ever growing catalog of GW observations is enabling deeper insights into the underlying source population. Readily accessible and reusable public data \cite{LIGOScientific:2025snk, Essick:2025zed, Ashton:2025xba} has lowered the barrier to enter for GW population studies and proliferated independent reanalyses, allowing new insights from GW data. However, the statistical significances of these findings are rarely assessed due to the technical and computational complexities involved in simulating realistic GW catalogs for mock population studies.

Following previous work \cite{Vitale:2025lms, Wolfe:2025yxu}, we demonstrated the feasibility of large-scale simulation and projection studies for GW population inference using full Bayesian PE for individual sources. In particular, we successfully repurposed previous PE for simulated sources drawn from other populations \cite{Vitale:2025lms, Wolfe:2025yxu} (see Appendix~\ref{section: Simulations}), providing a tractable route for robust and comprehensive GW population simulations without additional computationally expensive PE runs or relying on simplified prescriptions for mock PE. Though we may be able to conduct such studies using mock PE \cite{Fishbach:2019ckx, Farah:2023vsc} when the physics of interest is simple, like measuring a mass cutoff, this may not be true when probing the population in finer detail. However, we still neglected some physics like higher modes in the waveform model---which may impact population-level measurement of cuts in the mass spectrum \cite{Singh:2023aqh}---and made simplifying approximations for the detection of GW signals. A pertinent future data product could therefore be PE for simulated signals in real detector noise that were successfully recovered by the search pipelines used in LVK analyses, which could be repurposed for various target populations of interest as in this work. The development of machine-learning methods for population inference may also offer a promising avenue to efficiently build more significant backgrounds \cite{Leyde:2023iof, Jiang:2025jxt}.

Turning to our astrophysical results, we emphasized that interpretation of a mass gap inferred from GWTC-4 relies on prior assumptions about the source of GW231123 \cite{LIGOScientific:2025rsn} (the heaviest BH merger observed by the LVK), but that the presence of a feature at $\sim45M_\odot$ possibly associated with PISNe is collectively informed by the whole catalog. In our models (similar to Ref.~\cite{Tong:2025wpz}), this feature is the maximum secondary BH mass, but inferring the extrema of distributions can be prone to significant statistical fluctuations \cite{Mandel:2025qnh}. Our mock catalogs---simulated with the efficient procedure described above---revealed that the GWTC-4 constraints on this mass cutoff are compatible with expectations if the underlying population of secondary masses really was truncated at $45M_\odot$, and furthermore that spuriously inferring a cutoff as low as $\lesssim50M_\odot$ is unlikely if it is not present in reality. In light of this constraint on the lower edge of the PISN gap at 40--50$M_\odot$, it may be interesting to revisit the possible association of the peak in the distribution of BH masses at $\approx35M_\odot$ with pulsational PISNe, which was previously suggested as an unlikely origin \cite{Hendriks:2023yrw, Roy:2025ktr}.

On the other hand, the Bayes factor in favor of a mass cutoff in GWTC-4 was much stronger than for any of our simulated catalogs. One possible reason is that our simulated populations were limited to a maximum mass up to $90M_\odot$ (see Appendix~\ref{section: Simulations}); this is compatible with the GWTC-4 constraints for the \textsc{Single Power Law + 2 Peaks} (see Figure~\ref{figure: observations}) and \textsc{Broken Power Law + 2 Peaks} models, whereas the \textsc{Single Power Law + 2 Peaks + Cutoff} model prefers a maximum $m_1>90M_\odot$. Our simulations are therefore conservative, in the sense that they make a lower independent $m_2$ cutoff more difficult to distinguish from the global $m_1$ cutoff. There may also be features in the population indicative of the mass gap that we did not include in our models and simulations, e.g., mass--spin correlations \cite{Mould:2022ccw, Wang:2022gnx, Li:2023yyt, Pierra:2024fbl, Antonini:2024het, Tong:2025xir, Plunkett:2026pxt, Farah:2026jlc, Vijaykumar:2026zjy}. These features could impact the measurement of the mass gap, leading to tighter constraints from the real catalog than our simulations.

In contrast to Ref.~\cite{LIGOScientific:2025pvj}, from GWTC-4 we did not find preference for more subtle structure in the distribution of BH masses, such as breaks in the power-law continuum. Our projections for catalogs similar to that which may be available by the end of O4 suggested that these will continue to be poorly constrained relative to constraints on more clear-cut features such as the mass cutoff. Similarly, our simulations for the end of O4 did not result in especially improved GW-only constraints on the Hubble parameter. However, recent work suggests the application of more flexible population models to spectral-siren analyses may improve constraints \cite{Tong:2025xvd, Pierra:2026ffj, Tagliazucchi:2026gxn}, for which we plan to test the nonparametric \textsc{PixelPop} popultion model in future work.

With our larger simulated catalogs, constraints on the mass cutoff inferred from the population of BH masses alone continued to improve, with spurious measurements even less probable, implying confidence in the physical origin of the real feature will increase. However, if the mass cutoff indeed falls at 40--50$M_\odot$, constraining the uncertain $^{12}\mathrm{C}(\alpha,\gamma)^{16}\mathrm{O}$ reaction rate using current and near-future GW data alone is perhaps less promising than anticipated, as a large range of reaction rates all map to this same small range of maximum BH mass \cite{Farmer:2020xne, Mehta:2021fgz}. However, a sufficiently stringent lower bound alone may lead to tension with laboratory measurements or theoretical models \cite{deBoer:2017ldl, Shen:2023rco, Mukhamedzhanov:2025uvy, Hendriks:2023yrw, Wang:2025nhf}.

Ref.~\cite{Ray:2025xti} recently suggested that evidence for the PISN cutoff or gap is overestimated by parametric population models. Our analyses of GWTC-4 using \textsc{PixelPop} also did not find a strict mass cutoff, but rather a steeply declining merger rate. This, however, was consistent with analyses of catalogs simulated with a cutoff, though the inferred high-mass slopes tended to be steeper than found from the real catalog. It is possible the true population lies somewhere in between. Our nonparametric analyses also revealed that current parametric population models are likely mischaracterizing secondary-mass peaks at $\sim10M_\odot$ and $\sim30M_\odot$, and thus these models may need to be revised for future analyses. However, our nonparametric constraints did not indicate the presence of additional features in the distribution of BH masses $\sim60$--$70M_\odot$, as suggested by Refs.~\cite{MaganaHernandez:2024qkz, Wang:2025nhf, Pierra:2026ffj}.

We focused on inference of the lower-edge of the pair-instability mass gap, as the constraint on this lower cutoff from GWTC-4 is much more robust than a possible upper edge \cite{Tong:2025wpz}. Interesting future studies are therefore to: quantify the overall significance of a mass-gap detection from GW catalogs, including events below, above \cite{Ezquiaga:2020tns, Franciolini:2024vis}, and within the gap; test the results of correlated BH mass--spin models \cite{Antonini:2024het, Pierra:2024fbl}; and investigate whether origins directly tracing stellar collapse can be distinguished from other astrophysical formation processes.

\acknowledgments

We thank Fabio Antonini, Simona Miller, and Hui Tong for discussions.
M.M. is supported by a Research Fellowship from the Royal Commission for the Exhibition of 1851 and by the LIGO Laboratory through the National Science Foundation award No. PHY-2309200.
J.H., C.P., and N.E.W. are supported by the National Science Foundation Graduate Research Fellowship Program under Grant No. 2141064.
S.A-L. and S.V. are supported by the National Science Foundation Grant No. PHY-2045740.
The authors are grateful for computational resources provided by the LIGO Laboratory and supported by National Science Foundation Grants PHY-0757058 and PHY-0823459.
This research has made use of data or software obtained from the Gravitational Wave Open Science Center (\href{https://gwosc.org/}{gwosc.org}), a service of the LIGO Scientific Collaboration, the Virgo Collaboration, and KAGRA,
and is based upon work supported by NSF's LIGO Laboratory, which is a major facility fully funded by the National Science Foundation.

\appendix

\section{Models}
\label{section: Models}

We follow the implementation of the population likelihood and regularization thereof from Ref.~\cite{LIGOScientific:2025pvj}, including estimation of survey sensitivity \cite{Essick:2025zed, Tiwari:2017ndi} and use of PE samples \cite{LIGOScientific:2025snk, Ashton:2025xba}; we limit the relative Monte Carlo uncertainty of the likelihood estimator due to finite sampling to a maximum of unity \cite{Tiwari:2017ndi, Farr:2019rap, Essick:2022ojx, Talbot:2023pex, Heinzel:2025ogf}. Inference is performed using \textsc{gwax} \cite{matthew_mould_2025_15002875}, with \textsc{Dynesty} \cite{Speagle:2019ivv} for nested sampling \cite{Skilling:2004pqw, Skilling:2006gxv} through \textsc{Bilby} \cite{Ashton:2018jfp}. The mass models are summarized in Table~\ref{table: models}. The parameters of our models and the Bayesian priors we use for analyses of both real and simulated GW catalogs are given in Table~\ref{table: priors}.

\begin{table}
\caption{Parameters of the three parametric models we consider, with the priors chosen for inference and the values used for simulated populations. Prior intervals indicated the ranges for uniform distributions and ``Dir.'' indicates a flat symmetric Dirichlet distribution. The top section includes parameters governing redshift and BH spin distributions common to all parametric models. The second section includes BH mass parameters common to all models. The third contains the primary-mass power-law slope for the \textsc{Single Power Law + 2 Peaks} and \textsc{Single Power Law + 2 Peaks + Cutoff} models. The fourth contains the additional maximum secondary-mass parameters for the \textsc{Single Power Law + 2 Peaks + Cutoff} model, such that $m_2<\min(m_1,m_\mathrm{gap})$ with $m_\mathrm{gap}<m_\mathrm{max}$. The fifth section contains the parameters unique to the \textsc{Broken Power Law + 2 Peaks} model. The bottom section includes cosmological parameters for the spectral-siren population analyses.}
\centering
\begin{tabular}{cccc}
\hline\hline
Symbol & Description & Prior & Value  \\
\hline\hline
$\gamma$ & $z$ power law index & $[-10,10]$ & 3 \\
$\mu_a$ & Spin magnitude mode & $[0,1]$ & 0.3 \\
$\sigma_a$ & Spin magnitude width & $[0,1]$ & 0.2 \\
$\mu_\theta$ & Cosine spin tilt mode & $[-1,1]$ & 1 \\
$\sigma_\theta$ & Cosine spin tilt width & $[0,4]$ & 1.5 \\
\hline
$\mu_1$ & Low $m_1$ peak $[M_\odot]$ & $[5,15]$ & 10 \\
$\mu_2$ & High $m_1$ peak $[M_\odot]$ & $[25,50]$ & 34 \\
$\sigma_1$ & Low $m_1$ peak width $[M_\odot]$ & $[0,10]$ & 0.5 \\
$\sigma_2$ & High $m_1$ peak width $[M_\odot]$ & $[0,10]$ & 4 \\
$f_0$ & Fraction in $m_1$ power law & Dir. & 0.4 \\
$f_1$ & Fraction in low $m_1$ peak & Dir. & 0.5 \\
$f_2$ & Fraction in high $m_1$ peak & Dir. & 0.1 \\
$\delta_\mathrm{min}$ & Low-mass smoothing $[M_\odot]$ & $[0,10]$ & 5 \\
$m_\mathrm{min}$ & Minimum mass $[M_\odot]$ & $[3,10]$ & 5 \\
$m_\mathrm{max}$ & Maximum mass $[M_\odot]$ & $[30,150]$ & 90 \\
$\beta$ & $m_2$ power law index & $[-10,10]$ & 1 \\
\hline
$\alpha$ & $m_1$ power law index & $[-10,10]$ & $-2.5$ \\
\hline
$m_\mathrm{gap}$ & Maximum $m_2 \ [M_\odot]$ & $[30,150]$ & 45 \\
\hline
$\alpha$ & $m_1$ index below break & $[-10,10]$ & $-1.5$ \\
$\alpha_\mathrm{b}$ & $m_1$ index above break & $[-10,10]$ & $-4$ \\
$m_\mathrm{b}$ & $m_1$ break mass $[M_\odot]$ & $[25, 50]$ & 34 \\
\hline
$H_0$ & Hubble parameter [km\,s$^{-1}$\,Mpc$^{-1}$] & $[20,200]$ & 67.74 \\
$\Omega_\mathrm{m}$ & Matter density & $=0.3075$ & $0.3075$ \\
\hline\hline
\end{tabular}
\label{table: priors}
\end{table}

Though Ref.~\cite{LIGOScientific:2025pvj} now finds dimensionless BH spin magnitudes are better modeled as independently and identically distributed (IID) according to truncated Gaussians, as in Ref.~\cite{KAGRA:2021duu} and for reasons we specify in more detail in Appendix~\ref{section: Simulations}, we instead use IID nonsingular beta distributions \cite{Wysocki:2018mpo}. The polar spin tilt angles between the BH spins and binary orbital angular momenta are fit with IID truncated Gaussian distributions in cosine, with free location and scale parameters \cite{Vitale:2015tea, Talbot:2017yur, Vitale:2022dpa}, while the azimuthal angles are assumed to be distributed uniformly. The merger rate over is modeled as evolving over redshift $z$ according to a power law in $1+z$ \cite{Fishbach:2018edt}.

Ref.~\cite{LIGOScientific:2025pvj} used several models for the population of binary BH masses, taking as their default a distribution of primary BH masses $m_1$ with a broken power law and two Gaussian peaks, alongside a single power law for the distribution of binary mass ratios $0<q\leq1$. The primary and secondary mass ($m_2=qm_1$) distributions have independent minima and low-mass smoothing, but both have fixed maxima of $300M_\odot$. We employ different models. We assume that the $m_1$ and $m_2$ distributions have the same minimum and low-mass smoothing; we use a cubic smoothing function that results in population models with closed-form normalization constants (similar to Ref.~\cite{DeRenzis:2024dvx}) and normalize each component individually, such that the inferred branching fractions can be interpreted directly (unlike in Ref.~\cite{LIGOScientific:2025pvj}). We allow the maximum $m_1$ and $m_2$ to be free parameters that are either equal to each other or independent, the latter case allowing for a truncation in $m_2$ due to PISN that does not appear in the $m_1$ distribution, as found by Ref.~\cite{Tong:2025wpz} (note that this is the absolute maximum of $m_2$ and a given binary must also satisfy $m_2\leq m_1$). Alongside the two Gaussian peaks, we consider both single and broken power law models in $m_1$.

We also use \textsc{PixelPop} \cite{Heinzel:2024jlc} to model the merger rate as piecewise constant over fixed two-dimensional $m_1$--$m_2$ bins. We use 100 bins per dimensions logarithmically spaced between 3--150$M_\odot$. A log-normal prior is set for the rate in each bin and conditionally depends only on neighboring bins, thus enforcing a minimal smoothing prior that is much more efficient than generic Gaussian process priors \cite{Mandel:2016prl, Ray:2023upk, Callister:2023tgi, Farah:2024xub, Antonini:2025zzw}. This allows us to assess the measurability of population-level features for BH masses without making assumptions about which features are present a priori, as has been successfully demonstrated on real \cite{Heinzel:2024hva} and simulated GW data \cite{Alvarez-Lopez:2025ltt}.

\section{Simulations}
\label{section: Simulations}

The parameter values used for our simulated populations are given in Table~\ref{table: priors}. To simulate GW catalogs, we need to draw binary BHs with source properties $\theta$ from these populations, generate GW signals $s(\theta)$ added to detector noise $n$, and run PE on each event with data $d=n+s(\theta)$ that passes a detection threshold (``det'') to generate a set of posterior samples $\Theta$. Written probabilistically, the process of generating the catalog $\{\Theta_n,d_n,\theta_n\}_{n=1}^N$ of PE results for detected sources can be written \cite{Essick:2022ojx}
\begin{align}
p \left( \{ \Theta_n, d_n, \theta_n \} \right) =
\prod_{n=1}^N p(\Theta_n|d_n) P(\mathrm{det}|d_n) p(d_n|\theta_n) p(\theta_n)
\, ,
\label{equation: catalog}
\end{align}
where $p(\theta)$ is the target population, $p(d|\theta)$ is the GW likelihood \cite{Thrane:2018qnx, LIGOScientific:2025hdt, Talbot:2025vth}, $P(\mathrm{det}|d)$ is the (discrete) selection function, $p(\Theta|d) = \prod_i p(\Theta^i|d)$ is the PE posterior distribution corresponding to data $d$ as represented by a finite set of independent samples $\Theta = \{\Theta^i\}$, and $N$ is the number of detections in the catalog. The sequence of steps can be understood reading Eq.~\eqref{equation: catalog} right to left.

We follow the procedures in Refs.~\cite{Vitale:2025lms, Wolfe:2025yxu} to perform these simulations. In summary: GW signals are generated using the \textsc{IMRPhenomXP} waveform model \cite{Pratten:2020ceb}; Gaussian noise for the LIGO Hanford, LIGO Livingston, and Virgo detectors is generated according to their O4 design power spectral densities \cite{KAGRA:2013rdx, T2000012}; data containing signal and noise is classed as detected if the matched-filter signal-to-noise ratio (SNR) \cite{LIGOScientific:2025yae} is greater than 11 \cite{Essick:2023toz, Mould:2023eca} (note that this is technically an unphysical selection criterion \cite{Essick:2023upv} as the matched-filter SNR is computed using the true signal parameters, which in reality are not directly accessible from the data, and this also means that the population likelihood function is not self consistent, but in several previous works \cite{Heinzel:2024jlc, Mould:2025dts, Vitale:2025lms, Alvarez-Lopez:2025ltt, Wolfe:2025yxu} and indeed in the present work we found that this approximation does not lead to biased results in practice); for PE, we use the same waveform model and draw posterior samples using nested sampling \cite{Skilling:2004pqw, Skilling:2006gxv} as implemented in \textsc{Dynesty} \cite{Speagle:2019ivv} and \textsc{Bilby} \cite{Ashton:2018jfp}, with Bayesian priors set as in Ref.~\cite{Wolfe:2025yxu}; we estimate the survey sensitivity using the injection distribution from Ref.~\cite{Wolfe:2025yxu}.

\begin{figure*}
\centering
\includegraphics[width=1\textwidth]{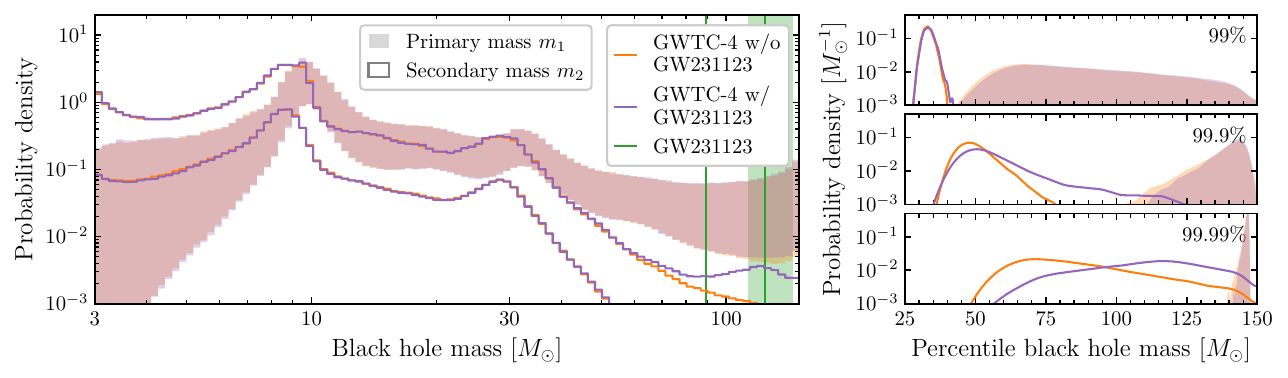}
\caption{Same as the bottom row of Figure~\ref{figure: observations}, now comparing \textsc{PixelPop} analyses of GWTC-4 with (purple) and without (orange) GW231123 included in the catalog.}
\label{figure: gw231123}
\end{figure*}

Even when using likelihood acceleration techniques \cite{Cornish:2010kf, Cornish:2021lje, Zackay:2018qdy, Leslie:2021ssu, Krishna:2023bug}, large-scale PE studies such as required here are very computationally expensive. We avoid having to do new PE runs and instead reuse the posterior samples produced in Refs.~\cite{Vitale:2025lms, Wolfe:2025yxu}, which each simulated 1600 GW events but from different populations, say, $q_1(\theta)$ and $q_2(\theta)$. We use a mixture $q(\theta) = f q_1(\theta) + (1 - f) q_2(\theta)$ as a proposal distribution from which we reweigh an already simulated catalog to our target populations of interest. The mixing fraction $f$ is chosen such that there is an equal number of detections from the two contributing populations as this maximizes the sample efficiency of reweighed catalogs, i.e.,
\begin{align}
f = \frac { q_2(\mathrm{det}) } { q_1(\mathrm{det}) + q_2(\mathrm{det}) }
\, ,
\end{align}
where
\begin{align}
q_i(\mathrm{det}) = \int \dd{d} \dd{\theta} P(\mathrm{det}|d) p(d|\theta) q_i(\theta)
\end{align}
is the fraction of detectable sources from the population $q_i(\theta)$. As the two population $q_1$ and $q_2$ from Refs.~\cite{Vitale:2025lms, Wolfe:2025yxu} are similar and the total number of detections considered in each are the same, we have $q_1(\mathrm{det}) \approx q_2(\mathrm{det})$ and consequently $f \approx 1-f \approx 0.5$.

Denoting by $q(\Theta,d,\theta) = p(\Theta|d) P(\mathrm{det}|d) p(d|\theta) q(\theta)$ the probabilistic model for a single event from the combined catalog $\{ \Theta_m, d_m, \theta_m \}_{m=1}^M$ from Refs.~\cite{Vitale:2025lms, Wolfe:2025yxu}, where $M>N$ is their total number of simulated events, we reweigh to our target populations by downselecting (without repetition) to $N$ events with each drawn according to
\begin{align}
p(\Theta,d,\theta) = q(\Theta,d,\theta) \frac{p(\theta)}{q(\theta)}
\, .
\end{align}
In words, for each event in the original simulated catalog $\{ \Theta_m, d_m, \theta_m \}_{m=1}^M$ drawn from population $q(\theta)$, we randomly select a subset of size $N$ according to probabilities $\propto p(\theta_m) / q(\theta_m)$. For the data-level posterior predictive checks in Section~\ref{section: Predictive checks}, the only additional step is selecting the highest-likelihood sample from $\Theta$ for each event.

We verified that the population posteriors inferred from these resampled catalogs are consistent with the simulated parameters listed in Table~\ref{table: priors}. Note, however, that this resampling procedure means each of our catalogs are bootstrapped realizations instead of being independent, i.e., some of the events included in one catalog may also be included another. For example, as we have $M=3200$ simulated events from Refs.~\cite{Vitale:2025lms, Wolfe:2025yxu} available, even without reweighing we would only be able to produce 10 completely independent catalogs of 300 events each, which is why we only consider 10 different bootstrap realizations for each of our analyses. Nonetheless, the apparent statistical variations (e.g., seen in Figure~\ref{figure: mgap}) demonstrate that our simulations indeed cover a range of plausible catalog realizations and results therefrom. Similarly, the relatively small number (3200) of simulated events with PE implies that the posterior spread in the highest-likelihood predictive test in Figure~\ref{figure: ppc} may be underestimated, but since the observed distribution is already consistent with the posterior predictions, a wider spread would only make it more so.

Also note that the supports of the target populations $p(\theta)$ must be covered by the support of the proposal population $q(\theta)$. As Refs.~\cite{Vitale:2025lms, Wolfe:2025yxu} simulated sources with spin magnitudes drawn from nonsingular beta distributions, which have no support at zero and maximal spins, the populations we consider must also respect this constraint and therefore we cannot simulate from truncated Gaussians as used in Ref.~\cite{LIGOScientific:2025pvj}, so we instead keep nonsingular beta distributions. Similarly, we cannot reweigh to spin population models from Refs.~\cite{Antonini:2024het, Antonini:2025zzw, Antonini:2025ilj} as the proposal spin magnitude distribution does not produce sufficiently many sources with large positive or negative effective spin. We are also limited by the maximum BH mass simulated in Refs.~\cite{Vitale:2025lms, Wolfe:2025yxu}, which is $\approx90M_\odot$.

\section{Impact of GW231123}
\label{section: Impact of GW231123}

In Figure~\ref{figure: gw231123}, we show the results when repeating the \textsc{PixelPop} analysis of GWTC-4 in Section~\ref{section: Nonparametric results} but now including GW231123. The inferred primary-mass distribution is essentially unchanged. The secondary-mass distribution has more support at higher masses where the lighter source BH of GW231123 is inferred to lie based on LVK PE results. The population 99\% $m_2$ mass when including GW231123 ($33_{-3}^{+4}M_\odot$) is almost unchanged from the result with GW231123 excluded ($33_{-3}^{+3}M_\odot$). The 99.9\% secondary mass $54_{-10}^{+46}M_\odot$ has substantially increased uncertainty from $49_{-7}^{+16}M_\odot$, while the 99.99\% value has similarly broad uncertainty as before but is shifted upward to $111_{-39}^{+30}M_\odot$ from $82_{-23}^{+48}M_\odot$.

\begin{figure*}
\centering
\includegraphics[width=1\textwidth]{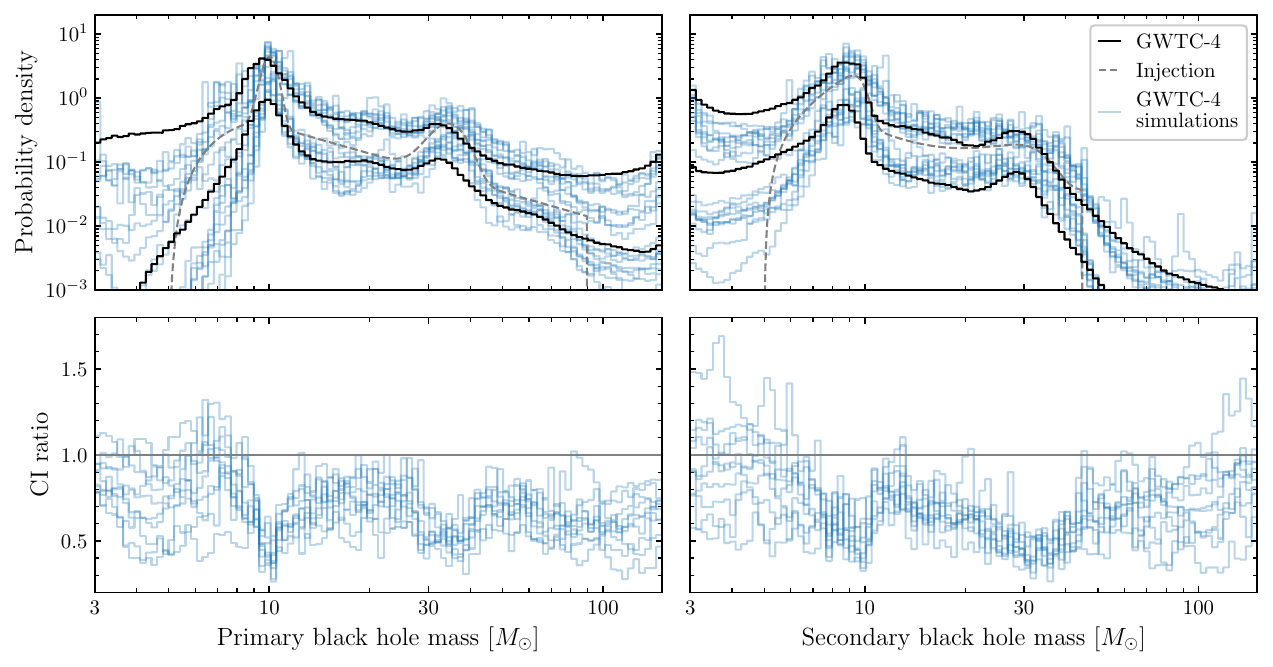}
\caption{Population distributions (top row) of primary (left column) and secondary (right column) BH masses in terms of the 90\% CI for each mass bin inferred by \textsc{PixelPop} from 10 catalogs (blue) of 150 events simulated from the \textsc{Single Power Law + 2 Peaks + Cutoff} population (dashed gray) without measurement uncertainty. GWTC-4 results (black) are included for reference. The ratios of the widths of each CI when single-event uncertainty is and is not included is shown for each mass bin in the bottom row, which ratios below unity (solid gray) indicating reduced population measurement uncertainty when single-event measurement uncertainties are neglected.}
\label{figure: deltas}
\end{figure*}

\section{Impact of single-event uncertainty}
\label{section: Impact of single-event uncertainty}

To test the impact of single-event measurement uncertainty on identifying a mass cutoff with a nonparametric population model, we repeat the \textsc{PixelPop} analyses of the 150-event catalogs from the \textsc{Single Power Law + 2 Peaks + Cutoff} population but assuming the true source parameters of each event are known exactly. The inferred mass distributions are shown in Figure~\ref{figure: deltas}. We also show the ratios between the 90\% CI widths of the merger rate in each mass bin when single-event uncertainty is and is not included.

Unsurprisingly, assuming point estimates for the source properties of each event tends to reduce uncertainty in the inferred population, particular at BH masses $\sim10M_\odot$ and 30--40$M_\odot$ coincident with identifiable features in the simulated population, where the widths of the 90\% CIs of the merger rate can be reduced by $\lesssim70\%$. For some catalogs however, the point estimates cause larger bin-to-bin fluctuations in the inferred merger rate, especially for secondary masses near the edges of parameter space where the 90\% CIs can widen by $\lesssim70\%$ relative to the same catalogs with single-event uncertainty included. Using again the 99.9\% percentile of the secondary-mass population as a proxy for the mass cutoff, we find its 90\% CI can be as narrow as $8M_\odot$, compared to the smallest width of $12M_\odot$ when single-event uncertainty is included (see Table~\ref{table: percentiles}). This suggests that it is not just the flexibility of the nonparametric \textsc{PixelPop} population model but also the measurement uncertainty inherent to GW data that impacts our ability to identify a possible mass cutoff with minimal model assumptions.

\bibliography{draft}

\end{document}